\documentclass[aps,showpacs,superscriptaddress,tightenlines,eqsecnum,nofootinbib]{revtex4}
\usepackage{graphicx,amssymb,amsbsy,bm,amsmath}
\usepackage{dsfont}
\usepackage{hyperref}

\newcommand{\vk}{\ensuremath{\vec{k}}}
\newcommand{\tmk}{\ensuremath{\theta_m(k)}}
\newcommand{\tm}{\ensuremath{\theta_m}}

\newcommand{\be}{\begin{equation}}
\newcommand{\ee}{\end{equation}}
\newcommand{\bea}{\begin{eqnarray}}
\newcommand{\eea}{\end{eqnarray}}

\begin{document}
\title{Production of a sterile species: quantum kinetics. }
\author{D. Boyanovsky}
\email{boyan@pitt.edu} \affiliation{Department of Physics and
Astronomy, University of Pittsburgh, Pittsburgh, Pennsylvania 15260,
USA}
\author{C. M. Ho} \email{cmho@phyast.pitt.edu}
\affiliation{Department of Physics and Astronomy, University of
Pittsburgh, Pittsburgh, Pennsylvania 15260, USA}
 \date{\today}
\begin{abstract}
Production of a sterile species
  is studied within an effective    model of  active-sterile neutrino
  mixing in a medium  in thermal equilibrium.
   The quantum kinetic equations for the distribution functions and coherences
  are obtained from two independent methods: the effective action  and the quantum master equation.
  The decoherence time
scale for active-sterile oscillations is $\tau_{dec} =
2/\Gamma_{aa}$, but the   evolution of the distribution functions is
determined
  by the two different time scales associated with the damping rates of the quasiparticle modes
  in the medium: $\Gamma_1=\Gamma_{aa}\cos^2\tm~;~\Gamma_2=\Gamma_{aa}\sin^2\tm$ where $\Gamma_{aa}$ is
  the interaction rate of the active species in absence of mixing and $\tm$ the mixing angle in the medium.
These  two time scales  are widely different away from MSW
resonances and preclude the kinetic description of active-sterile
production in terms of a simple rate equation.    We give the
complete set of quantum kinetic equations for the active and sterile
populations and coherences and discuss in detail the various
approximations.  A generalization of the active-sterile transition
probability \emph{in a medium} is provided via the quantum master
equation. We  derive explicitly the usual quantum kinetic equations
in terms of the ``polarization vector'' and show their equivalence
to those obtained from the quantum master equation and effective
action.

\end{abstract}


\maketitle

\section{Introduction}\label{sec:intro}
Sterile neutrinos, namely weak interaction singlets,  are acquiring
renewed attention as potential candidates for cold or warm dark
matter\cite{dodelson,asaka,shi,kev1,hansen,kev2,kev3,kusenko,kou,dolgovrev,pastor,hannestad,biermann,michaDM},
and may also be relevant in stellar collapse\cite{raffeltSN,fuller},
primordial nucleosynthesis\cite{fuller2,fuller3}, and as potential
explanation of the anomalous velocity distributions of
pulsars\cite{segre,fullkus,kuse2}. Although sterile neutrinos are
ubiquitous in extensions of the standard
model\cite{book1,book2,book3,raffelt}, the MiniBooNE
collaboration\cite{miniboone} has recently reported   results in
contradiction with   those from LSND\cite{lsnd1,lsnd2} that
suggested a   sterile neutrino with $\Delta m^2 \sim
1~\textrm{eV}^2$ scale. Although the MiniBooNE results  hint at an
excess of events below $475~\mathrm{MeV}$ the analysis distinctly
excludes two neutrino appearance-only from $\nu_\mu \rightarrow
\nu_e$ oscillations  with a mass scale $\Delta m^2\sim
1~\textrm{eV}^2$, perhaps ruling out a  \emph{light} sterile
neutrino. However, a recent analysis\cite{malto} suggests that while
  $(3+1)$ schemes are strongly disfavoured, $(3+2)$ neutrino schemes
  provide a good fit to both the LSND and MiniBooNE data, including the low energy
  events, because of the possibility of CP violation in these schemes, although
  significant tension remains.

However, sterile neutrinos as dark matter candidates would require
masses in the $\mathrm{keV}$
range\cite{dodelson,asaka,shi,kev1,hansen,kev2,kev3,kou,pastor,hannestad},
hence the MiniBooNE result does not  constrain a heavier variety of
sterile neutrinos.   The  radiative decay of $\mathrm{keV}$
neutrinos would contribute to the X-ray
background\cite{hansen,Xray}. Analysis from the X-ray background in
clusters provide constraints on the masses and mixing angles of
sterile neutrinos\cite{kou,boyarsky,hansen2,kou2}, and recently it
has been suggested that precision laboratory experiments on $\beta$
decay in tritium   may be sensitive to $\sim \textrm{keV}$
neutrinos\cite{shapolast}. Being weak interaction singlets, sterile
neutrinos can only be produced via their mixing with an active
species, hence any assessment of the possibility of sterile
neutrinos as dark matter candidates or their role in supernovae must
begin with understanding their production mechanism. Pioneering work
on the   description of neutrino oscillations and decoherence in a
medium was cast in terms of kinetic equations for a flavor ``matrix
of densities''\cite{dolgov} or in terms of $2\times 2$ Bloch-type
equations for flavor quantum mechanical
states\cite{stodolsky,enquist}. A general field theoretical approach
to neutrino mixing and kinetics was presented in \cite{sigl,raffkin}
(see also \cite{raffelt}), however, while such   approach in
principle yields the time evolution of the distribution functions,
sterile neutrino production in the early Universe is mostly studied
in terms of simple phenomenological rate
equations\cite{dodelson,kev1,cline,kainu,foot,dibari}. An early
  approach\cite{cline} relied on a Wigner-Weisskopf effective
Hamiltonian for the quantum mechanical states in the medium, while
numerical studies of sterile neutrinos as possible dark matter
candidates\cite{kev1,dibari} rely on an approximate approach which
inputs an effective production rate in terms of a time averaged
transition probability\cite{kainu,foot}. More recently the sterile
production rate \emph{near an MSW resonance} including hadronic
contributions  has been studied in ref.\cite{shapo}.

The rich and complex dynamics of oscillations, decoherence and
damping is of \emph{fundamental} and phenomenological importance not
only in neutrino cosmology   but also in the dynamics of meson
mixing and CP violation\cite{cp,beuthe}. In ref.\cite{dolokun} it
was argued that the spinor nature of neutrinos is not relevant to
describe the dynamics of mixing and oscillations at high energy
which can then be studied within a (simpler) quantum field theory of
meson degrees of freedom.

Recently we reported on a  study\cite{hobos} of mixing, decoherence
and relaxation in a theory of mesons which provides an  accurate
  description of similar phenomena for mixed neutrinos. This
effective theory incorporates interactions that model the medium
effects associated with charge and neutral currents for neutrinos
and yield a robust picture of the non-equilibrium dynamics of
mixing,   decoherence and equilibration which is remarkably general.
The fermion nature of the distributions and Pauli blocking effects
can be simply accounted for in the final result\cite{hobos}.  This
study implemented quantum field theory methods   to obtain the
non-equilibrium effective action for the ``neutrino'' degrees of
freedom. The main ingredient in the time evolution is the \emph{full
propagator} for the ``neutrino'' degrees of freedom in the medium.
The complex poles of the propagator yield the dispersion relation
and damping rates of quasiparticle modes in the medium. The
dispersion relations are found to be the usual ones for neutrinos in
a medium with the index of refraction correction from forward
scattering. For the case of two flavors, there are \emph{two damping
rates} which are widely different away from MSW resonances. The
results of this study motivated\cite{hozeno} a deeper scrutiny of
the rate equation which is often used to study sterile neutrino
production in the early Universe\cite{foot,kev1,dibari}.

One of the observations in\cite{hozeno} is that  the emergence of
\emph{two widely different} damping time scales precludes a reliable
kinetic description in terms of a \emph{time averaged transition
probability} suggesting that a simple rate equation to describe
sterile neutrino production in the early Universe far away from MSW
resonances may not be reliable.

{\bf Motivation and goals:} The broad potential relevance of sterile
neutrinos as warm dark matter candidates in cosmology and their
impact in the late stages of stellar collapse warrant a deeper
scrutiny of the quantum kinetics of production of the sterile
species. Our goal is to provide a quantum field theory study of the
non-equilibrium dynamics of mixing, decoherence and damping and to
obtain the quantum kinetic equations that determine the production
of a sterile species. We make progress towards this goal within a
meson model with one active and one sterile degrees of freedom
coupled to a bath of mesons in equilibrium discussed in
ref.\cite{hobos}. As demonstrated by the results of ref.\cite{hobos}
this (simpler) theory provides a remarkable effective description of
propagation,  mixing, decoherence and damping of neutrinos in a
medium. While  ref.\cite{hobos} studied the approach to equilibrium
focusing on the one body density matrix  and  single quasiparticle
dynamics, in this article we obtain the non-equilibrium effective
action, the quantum master equation and the complete set of quantum
kinetic equations for the distribution functions and coherences. We
also establish a generalization of the active-sterile transition
probability based on the quantum master equation. In distinction
with a recent quantum field theory treatment\cite{shapo} we seek to
understand the quantum kinetics of production not only near MSW
resonances, at which both time scales concide\cite{hobos,hozeno} but
far away from the resonance region where the damping time scales are
widely separated\cite{hobos,hozeno}.

{\bf Similarities and differences:} The \emph{scalar field model}
that we study has many similarities with the neutrino case but also
important differences. {\bf Similarities:} as demonstrated in our
previous study\cite{hobos} a) the scalar model describes ``flavor
mixing'' in a similar manner as in the case of neutrinos, where
mixing arises from off-diagonal mass matrix elements, b) a medium
induced ``matter potential'' which arises from the
forward-scattering contribution to the (self) energy, c) the
dispersion relation for the propagating modes is \emph{identical} to
those of neutrinos in a medium\cite{notzold,hozeno}, d) the
effective mixing angles in the medium have a functional form
\emph{identical} to those for neutrinos in a
medium\cite{raffelt,notzold}, e) the form of the transition
probability for ensemble averages in the medium is \emph{identical}
to that for the active-sterile neutrino transition
probability\cite{hozeno}, f) the relationship between the damping
rates of the propagating modes and the active collision rate is
identical to the neutrino case\cite{hozeno} and g) as shown in
detail in section (\ref{sec:polar}) the kinetic equations obtained
are \emph{identical} to those in terms of the polarization vector
often quoted in the neutrino literature (see section
(\ref{sec:polar})). {\bf Differences:} There are obvious differences
with the neutrino case that should not be overlooked: a) spinor and
chirality structure: although this is a clear difference, it is
important to highlight that neither the quantum mechanical
description of neutrino mixing nor the phenomenological description
of neutrino kinetics account for either spinorial structure or
chirality. b) Fermionic vs. Bosonic degrees of freedom, the most
obvious difference is in the distribution functions, however the
results obtained in sections (\ref{sec:QK}-\ref{sec:polar}) for the
kinetic description allow a straightforward replacement of the
distribution functions for the Fermi-Dirac expressions thus
automatically including Pauli blocking. c) An important difference
is the matter potential, in the scalar model this is given by the
one-loop Hartree self-energy, which is manifestly positive, whereas
in the case of neutrinos the matter potential features a CP-odd and
a CP-even contribution\cite{notzold} and it can feature either sign.
The existence of an MSW resonance hinges on the sign of the self
energy, in particular on the CP-odd component. However, this
important difference notwithstanding, our study does not rely on or
require  a specific form of the matter potential, \emph{only the
fact that the matter potential is diagonal in the flavor basis and
in the case under consideration only the active-active matrix
element is non-vanishing}. Whether or not there is an MSW resonance
depends on the specific form of the matter potential and in the case
of neutrinos, on the CP-odd (lepton and baryon asymmetry) component
of the background. Our study addresses all possible cases quite
generally without the need to specify the sign (or any other
quantitative aspect) of the matter potential.

{\bf Summary of results:}
\begin{itemize} \item  {\bf i:}  We obtain the quantum kinetic
equations for production by two different but complementary methods:

{\bf a):} the non-equilibrium effective action obtained by
integrating out the ``bath degrees of freedom''. This method
provides a non-perturbative  Dyson-like resummation of the
self-energy radiative corrections  and leads to the full propagators
in the medium.   This method makes explicit that the ``neutrino''
propagator in the medium along with the generalized
fluctuation-dissipation relation of the bath in equilibrium are the
essential ingredients for the kinetic equations and allows to
  identify the various approximations. It
unambiguously reveals the emergence of \emph{two relaxation time
scales} associated with the damping rates of the propagating modes
in the medium $\Gamma_1=
\Gamma_{aa}\cos^2\tm~;~\Gamma_2=\Gamma_{aa}\sin^2\tm$ where
$\Gamma_{aa}$ is the interaction rate of the active species and
$\tm$ the mixing angle in the  medium, confirming the results of
references\cite{hobos,hozeno}. These time scales determine the
kinetic evolution of the distribution functions and coherences.

 {\bf b):} the quantum master equation for the \emph{reduced} density
matrix, which is obtained by \emph{including the lowest order medium
corrections to the dispersion relations (index of refraction) and
mixing angles into the unperturbed Hamiltonian}. This method
automatically builds in the correct propagation frequencies and
mixing angles in the medium.

From the quantum master equation we obtain the   kinetic equations
for the distribution functions and coherences. These are identical
to those obtained with the non-equilibrium effective action to
leading order in perturbative quantities. After discussing the
various approximations and their regime of validity we provide the
full set of quantum kinetic equations for the active and sterile
production as well as coherences. These are given by equations
(\ref{dotn11fin}-\ref{Nsat}) in a form amenable to numerical
implementation.  We show that if the initial density matrix is
off-diagonal in the basis of the propagating modes in the medium,
the off-diagonal coherences are damped out in a decoherence time
scale $\tau_{dec} = 2/\Gamma_{aa}$. The damping of these off
diagonal coherences leads to an equilibrium reduced density matrix
\emph{diagonal in the basis of propagating modes in the medium}.

\item{\bf  ii:} We elucidate the nature of the various approximations
that lead to the final set of quantum kinetic equations and discuss
the   interplay between oscillations, decoherence and damping within
the realm of validity of the perturbative expansion.

\item {\bf iii:}  We  introduce a generalization of the active-sterile
transition probability \emph{in the medium} directly based on the
quantum density matrix approach.  The transition probability depends
on both time scales $1/\Gamma_1,1/\Gamma_2$, and the oscillatory
term arising from the   interference of the $1,2$ modes in the
medium is damped out on the decoherence time scale $\tau_{dec}$ but
this is \emph{not} the relevant time scale for the build up of the
populations or the transition probability far away from an MSW
resonance.

\item {\bf iv:} We derive the quantum kinetic equation for the
``polarization vector'' often used in the literature \emph{directly}
from the  kinetic equations obtained from the quantum master
equation under a clearly stated approximation. We argue that the
  kinetic equations   obtained from the quantum master equation exhibit more
clearly   the time scales for production and decoherence and reduce
to a simple set within the regime of reliability of perturbation
theory.  We discuss the shortcomings of the phenomenological rate
equation often used in the literature for numerical studies of
sterile neutrino production.

\end{itemize}

In section (\ref{sec:model}) we introduce the model, obtain the
effective action, and the full propagator from which we extract the
dispersion relations and damping rates. In section (\ref{sec:QK}) we
define the active and sterile distribution functions and obtain
their quantum kinetic non-equilibrium evolution from the effective
action, discussing the various approximations. In section
(\ref{sec:QME}) we obtain the quantum Master equation for the
reduced density matrix, also discussing the various approximations.
In this section we obtain the full set of quantum kinetic equations
for the populations and coherences and show their equivalence to the
results from the effective action. In section (\ref{sec:transprob})
we study the kinetic evolution of the off-diagonal coherences and
introduce a generalization of the active-sterile transition
probability \emph{in a medium} directly from the quantum master
equation. In section (\ref{sec:polar}) we establish the equivalence
between the kinetic equations obtained from the quantum master
equation and those   most often used in the literature in terms of a
``polarization vector'', along the way identifying the components of
this ``polarization vector'' in terms of the populations of the
propagating states in the medium $1,2$ and the coherences. While
this formulation is equivalent to the quantum kinetic equations
obtained from the master equation and effective action, we argue
that the latter formulations yield more information, making explicit
that the \emph{fundamental} damping scales are the widths of the
quasiparticle modes in the medium and allow to define the
generalization of the transition probability \emph{in the medium}.
We also discuss the shortcomings of the phenomenological rate
equations often invoked for numerical studies of sterile neutrino
production. Section (\ref{sec:conclu}) summarizes our conclusions.
Two appendices elaborate on  technical aspects.

\section{The model, effective action,   and distribution
functions. }\label{sec:model}

 We consider a model of mesons
with two flavors $a ,s$ in interaction with a `` vector boson'' and
a ``flavor lepton'' here denoted as  $W$, $\chi_a$ respectively,
modeling charged and neutral current interactions in the standard
model. This model has been proposed as an effective description of
neutrino mixing, decoherence and damping in a medium   in
ref.\cite{hobos} to which we refer reader   for details. As it will
become clear below, the detailed nature of the bath fields
$W,\chi_a$ is only relevant through their equilibrium correlation
functions which can be written in dispersive form.

In terms of the field doublet

\be \Phi =  \Bigg( \begin{array}{c}
       \phi_a \\
              \phi_s \\
            \end{array} \Bigg)   \label{doublets}\ee the Lagrangian
            density is
\be \mathcal{L}  =  \frac{1}{2} \left\{ \partial_{\mu} \Phi^T
\partial^{\mu} \Phi -   \Phi^T  \mathds{M}^2    \Phi \right\}+  \mathcal{L}_0[W,\chi]+G \,W\,\phi_a
\chi_a
 +G\phi^2_a \chi^2_a   \label{lagra} \ee where the mass matrix
 is given by

\be \mathds{M}^2  = \left( \begin{array}{cc}
                                   M^2_{aa} & M^2_{as} \\
                                   M^2_{as} & M^2_{ss} \\
                                 \end{array} \right)
                                 \label{massmatrices} \ee and
 $\mathcal{L}_0[W,\chi]$ is the free field Lagrangian density for $W,\chi$
 which need not be specified.

 The mesons $\phi_{a,s}$ play the
 role of the active and sterile flavor  neutrinos, $\chi_{a }$ the role of the
 charged lepton associated with the active flavor and $W$   a  charged  current, for example
 the proton-neutron current
 $\overline{p}\gamma^\mu(1-g_A\gamma_5)n$ or a similar quark
 current. The coupling $G$ plays the role of $G_F$. The interaction between the ``neutrino''
  doublet and the $W,\chi_a$ fields  is of the same form as that studied in ref.\cite{raffelt,sigl,raffkin}
   for neutral and charged current interactions.

   The last term in the Lagrangian density (\ref{lagra}) allows to
   model the matter effective potential from forward scattering in
   the medium by replacing $\chi^2_a$ by its expectation value in
   the statistical ensemble, $\langle \chi^2_a \rangle$. The
   resulting term $G\phi^2_a \langle \chi^2_a \rangle$ effectively
   models a matter potential from forward scattering in the medium\cite{raffelt}.
   While in the bosonic case $\langle \chi^2_a \rangle$ is
   manifestly positive, in the fermionic case the effective
   potential from forward scattering in the  medium features two
   distinct contributions\cite{notzold}: a CP odd contribution which is
   proportional to the lepton and baryon asymmetries, and a CP even
   contribution that only depends on the temperature. However, as it will become clear below,
    we do not need to specify the
 precise form of the matter potential or of the bath degrees of freedom, only the fact that the
 matter potential is diagonal in the flavor basis with only entry in the $a-a$ component, and the
 spectral properties of the  correlation
 function of  bath degrees of freedom  are necessary.

The flavor $\phi_{a,s}$ and the mass basis fields $\varphi_{1,2}$
are related  by an orthogonal transformation $\Phi = U(\theta)\,
\varphi$

\be   \left(\begin{array}{c}
      \phi_a \\
      \phi_s\\
    \end{array}\right) =  U(\theta) ~\Bigg(\begin{array}{c}
                                               \varphi_1 \\
                                               \varphi_2\\
                                             \end{array}\Bigg)~~;~~U(\theta)
                                             = \Bigg( \begin{array}{cc}
                           \cos\theta & \sin\theta \\
                           -\sin\theta & \cos\theta \\
                         \end{array} \Bigg) \label{trafo} \ee where
the orthogonal matrix $U(\theta)$ diagonalizes the mass matrix
$\mathds{M}^2$, namely

\be U^{-1}(\theta)\,\mathds{M}^2 \, U(\theta) = \Bigg(
\begin{array}{cc}
                                         M^2_1 &0 \\
                                         0 & M^2_2 \\
                                       \end{array} \Bigg)
                                       \label{diagM} \ee

In the flavor basis the mass matrix $\mathbb{M} $ can be written in
terms of the vacuum mixing angle $\theta$ and the eigenvalues of the
mass matrix as

\be \mathds{M}^2  = \overline{M}^{\,2}\,\mathds{1}+\frac{\delta
M^2}{2} \left(\begin{array}{cc}
                                                                -\cos 2\theta & \sin2\theta \\
                                                                \sin 2\theta & \cos 2\theta \\
                                                              \end{array}
\right) \label{massmatx2}\ee where we introduced

\be \overline{M}^{\,2} =\frac{1}{2}(M^2_1+M^2_2)~~;~~ \delta M^2 =
M^2_2-M^2_1 \label{MbarDelM}\,. \ee

For the situation under consideration with $\textrm{keV}$ sterile
neutrinos with small vacuum mixing angle $\theta \ll 1$ \be M_{aa}
\sim M_1~;~M_{ss} \sim M_2 \label{masses} \ee and in the vacuum
\be\phi_a \sim \phi_1~;~\phi_s \sim \phi_2\,.\ee  We focus on the
description of the dynamics of the ``system fields'' $\phi_\alpha,~
\alpha=a,s$. The strategy is to consider the time evolved full
density matrix and trace over the bath degrees of freedom $\chi,W$.
It is convenient to write the Lagrangian density (\ref{lagra}) as
\begin{equation}\label{lagra2}
{\cal L}[\phi_\alpha,\chi_\alpha,W]= {\cal L}_{0}[\phi]+{\cal
L}_{0}[W,\chi]+G\phi_a \mathcal{O}_a+ G \phi^2_a \chi^2_a
\end{equation}  where \be \mathcal{O}_a =   \chi_a \,W \,.\label{calO}\ee
and ${\mathcal L}_{0}[\cdots]$  are the free  Lagrangian densities
for the   fields $\phi_\alpha,\chi_a,W$ respectively.  The fields
$\phi_\alpha$ are considered as the ``system''  and the fields
$\chi_a,W$ are treated as a bath in thermal equilibrium at a
temperature $T\equiv 1/\beta$. We consider a factorized initial
density matrix at a time $t_0 =0$ of the form
\begin{equation}
\widehat{\rho}(0) =  {\rho}_{\Phi}(0) \otimes
\rho_{B}(0)~~;~~\rho_{B}(0)= e^{-\beta\,H_0[\chi,W]}
\label{inidensmtx}
\end{equation} where $H_0[\chi,W]$ is   Hamiltonian for
the fields $\chi_a,W$ in absence of interactions with the neutrino
field $\phi_a$.

Although this factorized form of the initial density matrix leads to
initial transient dynamics, we are interested in the long time
dynamics, in particular in the   long time limit.

 The  bath fields $\chi_\alpha,W$  will be ``integrated
out'' yielding a reduced density matrix for the fields $\phi_\alpha$
in terms of an effective real-time functional, known as the
influence functional\cite{feyver} in the theory of quantum brownian
motion. The reduced density matrix can be represented by a path
integral in terms of the non-equilibrium effective action that
includes the influence functional.  This method has been used
extensively to study quantum brownian motion\cite{feyver,leggett},
and quantum kinetics\cite{boyalamo,hoboydavey} and more recently in
the study of the non-equilibrium dynamics of thermalization in a
similar model\cite{hobos}. The time evolution of the initial density
matrix is given by

\begin{equation}\label{rhooft}
\widehat{\rho}(t )= e^{-iH(t -t_0)}\widehat{\rho}(t_0)e^{iH(t -t_0)}
\end{equation} Where the total Hamiltonian $H$ is
\begin{equation}\label{hami}
H=H_{0}[\phi] + H_{0}[\chi,W]+H_I[\phi,\chi,W]
\end{equation} Denoting all the fields collectively as $X$ to simplify notation, the
density matrix elements in the field basis are given by \be \langle
X|\widehat{\rho}(t)|X'\rangle = \int DX_i  DX'_i~ \langle X|
e^{-iH(t -t_0)}|X_i\rangle~\langle
X_i|\widehat{\rho}(t_0)|X'_i\rangle~\langle X'_i| e^{iH(t
-t_0)}|X'\rangle \,. \label{denmt} \ee The density matrix elements
in the field basis can be expressed as a path integral by using the
representations \be \langle X |e^{-iH(t -t_0)}|X_i\rangle = \int
\mathcal{D}X^+ ~ e^{i\int_{t_0}^{t }dt \int d^3x
\mathcal{L}[X^+]}~~;~~X^+(t_0)=X_i;X^+(t )=X \,. \label{PIplus}\ee
Similarly \be \langle X'_i|e^{iH(t -t_0)}|X' \rangle = \int
\mathcal{D}X^- ~ e^{-i\int_{t_0}^{t }dt\int d^3x
\mathcal{L}[X^-]}~~;~~X^-(t_0)=X'_i;X^-(t )=X' \,. \ee Therefore the
full time evolution of the density matrix can be systematically
studied via the path integral \be \mathcal{Z}=\int \mathcal{D}X^+
\mathcal{D}X^- e^{i\int_{t_0}^{t }dt \int d^3x
\left\{\mathcal{L}[X^+]-\mathcal{L}[X^-]\right\}} \,,
\label{PItot}\ee with the boundary conditions discussed above. This
representation allows to obtain expectation values or  correlation
functions $\mathcal{C}(X,X_i,X'_i,X';t,t')$ which depend on the
values of the fields $X_{i};X'_{i}$ through the initial conditions.
In order to obtain expectation values or correlation functions in
the full time evolved density matrix, the results from the path
integral must be averaged in the initial density matrix
$\widehat{\rho}(t_0)$, namely \be \langle \mathcal{C}(X,X';t,t')
\rangle \equiv \int DX_i DX'_i~ \langle
X_i|\widehat{\rho}(t_0)|X'_i\rangle~ \mathcal{C}(X,X_i,X'_i,X';t,t')
\label{corre}\,.\ee

\begin{figure}[h!]
\begin{center}
\includegraphics[width=2.5in,keepaspectratio=true]{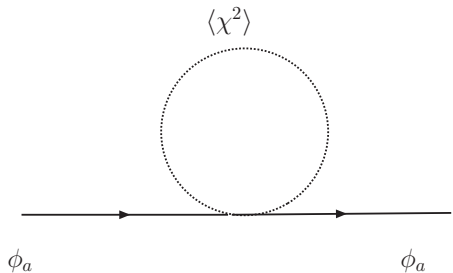}
\caption{One loop self-energy for the active species at order $G$,
corresponding to the matter potential $V_{aa}=G \langle \chi^2
\rangle$. } \label{fig:vacpot}
\end{center}
\end{figure}

We will only study  correlation functions of the  ``system'' fields
$\phi_\alpha$, therefore we carry out the trace over the $\chi_a$
and $W$ degrees of freedom   in the path integral (\ref{PItot})
systematically in a perturbative expansion in $G$. The resulting
series is re-exponentiated to yield the \emph{non-equilibrium}
effective action and the generating functional of connected
correlation functions of the fields $\phi_\alpha$. This procedure
has been explained in detail in references\cite{boyalamo,hoboydavey}
and more recently in \cite{hobos} within a model similar to the one
under consideration. Following the procedure detailed in these
references we obtain the non-equilibrium effective action up to
order $G^2$ and quadratic in the fields $\phi_\alpha$ neglecting
higher order non-linearities,
\begin{eqnarray}\label{influfunc} iL_{eff}[\phi^+,\phi^-] & = &
\sum_{\vec k}\Bigg\{ \frac{i}{2} \int dt
\Big[\dot{\phi}^+_{\alpha,\vec k}(t)\dot{\phi}^+_{\alpha,-\vec
k}(t)- \phi^+_{\alpha,\vec k}(t)(k^2\mathds{1}+ \mathds{M}^2 +\mathds{V} ) \phi^+_{\beta,-\vec k}(t)  \nonumber \\
& &
 -\dot{\phi}^-_{\alpha,\vec k}(t)\dot{\phi}^-_{\alpha,-\vec k}(t)+\phi^-_{\alpha,\vec k}(t)(k^2\,\mathds{1}
 + \mathds{M}^2 +\mathds{V} )
  \phi^-_{-\vec k}(t) \Big]   \nonumber \\
&   &  - \frac{G^2}{2} \int  dt \int  dt' \left[ \phi^+_{a,\vec
k}(t) {\cal G}^{++}(k;t,t')\phi^+_{a,-\vec k}(t')+ \phi^-_{a,\vec
k}(t){\cal G}^{--} (k;t,t') \phi^-_{a,-\vec k}(t') \right. \nonumber \\
&& \left. -\phi^+_{a,\vec k}(t){\cal G}^{+-} (k;t,t')\phi^-_{a,-\vec
k}(t')- \phi^-_{a,\vec k}(t){\cal G}^{-+} (k;t,t')\phi^+_{a,-\vec
k}(t')\right] \Bigg\}
\end{eqnarray} where the \emph{matter potential} is \be \mathds{V} = \left(%
\begin{array}{cc}
  V_{aa} & 0 \\
  0 & 0 \\
\end{array}%
\right)~~;~~ V_{aa} = G \langle \chi^2_a \rangle \label{matV}\,,\ee
with the average in the initial bath density matrix. In the bosonic
model, the corresponding one-loop diagram at order $G$ that yields
the \emph{matter potential} and effectively models forward
scattering in the medium is depicted in figure (\ref{fig:vacpot}).

In the fermionic theory, the matter potential in a medium at finite
temperature and density has two distinct
contributions\cite{notzold}: a CP-odd term proportional to the
lepton and baryon asymmetries and a CP-even term that only depends
on the temperature. The sign of these contributions may be either
positive or negative depending on which term
dominates\cite{notzold}. The presence of an MSW resonance in the
medium depends crucially on the CP-odd contribution. In the case of
sterile neutrinos with masses in the $keV$ range, only for
non-vanishing lepton asymmetry is there an MSW resonance. However,
the \emph{only} important point for the analysis that follows is
that the matter potential is diagonal in the flavor basis, with only
entry $V_{aa}$, namely the \emph{form} of the matrix given by eqn.
(\ref{matV}), but of course the matrix element $V_{aa}$ itself will
be different for fermions.

The correlation functions $\mathcal{G}(t,t') \sim \langle
\mathcal{O}_a(t) \mathcal{O}_a(t') \rangle = \langle W(t)
W(t')\rangle\,\langle\chi_a(t)\chi_a(t')\rangle$ are also determined
by averages in the initial equilibrium bath density matrix and their
explicit form is given in reference\cite{hobos} (see also appendix
(\ref{app:qme})).

Performing the trace over the bath degrees of freedom the resulting
non-equilibrium effective action acquires a simpler form in terms of
the Wigner center of mass and relative
variables\cite{hobos,boyalamo,hoboydavey} \be \Psi_\alpha(\vec x,t)
=   \frac{1}{2} \left(\phi^+_\alpha(\vec x,t) + \phi^-_\alpha(\vec
x,t) \right) \; \; ; \; \; R_\alpha(\vec x,t) =
\left(\phi^+_\alpha(\vec x,t) - \phi^-_\alpha(\vec x,t)
\right)~~;~~\alpha=a,s \label{wigvars} \ee and a corresponding
Wigner transform of the initial density matrix for the $\phi$
fields. See ref.\cite{hobos} for details. The resulting form allows
to cast the dynamics of the Wigner center of mass variable as a
\emph{stochastic} Langevin functional equation, where the effects of
the bath enter through a dissipative kernel and a stochastic noise
term, whose correlations obey a generalized fluctuation-dissipation
relation\cite{hobos,boyalamo,hoboydavey}. In terms of spatial
Fourier transforms the time evolution of the center of mass Wigner
field $\Psi$ is given by the following Langevin (stochastic)
equation (see derivations and details in
refs.\cite{feyver,leggett,hobos,boyalamo,hoboydavey})
\begin{eqnarray}
&& \ddot{\Psi}_{\alpha,\vec k}(t)+(k^2\,\delta_{\alpha
\beta}+\mathbb{M}^2_{\alpha
\beta}+\mathds{V}_{\alpha\beta})\Psi_{\beta,\vec k}(t)+\int_0^t dt'
~ \Sigma_{\alpha \beta}(k;t-t')
\Psi_{\beta,\vec k}(t')=\xi_{\alpha,\vec k}(t) \nonumber \\
&& \Psi_{\alpha,\vec k}(t=0)= \Psi^0_{\alpha,\vec k}~~ ; ~~
\dot{\Psi}_{\alpha,\vec k}(t=0)= \Pi^0_{\alpha,\vec k}
\label{langevin}
\end{eqnarray} where $ \Psi^0_{\alpha,\vec k},\Pi^0_{\alpha,\vec k}$
are the initial values of the field and its canonical momentum. The
matter potential $\mathds{V}$ in the equation of motion
(\ref{langevin}) effectively \emph{models} the general form of the
matter potential in the fermionic case. The specific value and sign
of $V_{aa}$ is not relevant for the general arguments presented
below.

\begin{figure}[h!]
\begin{center}
\includegraphics[width=2.5in,keepaspectratio=true]{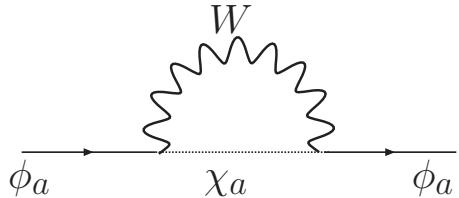}
\caption{One loop self-energy for the active species to order $G^2$.
The cut discontinuity across the $W-\chi$ lines yields the imaginary
part $\mathrm{Im}\widetilde{\Sigma}_{aa}(k;\omega)$.
 } \label{fig:loop}
\end{center}
\end{figure}

The stochastic noise $\xi_{\alpha,\vec k}(t)$ is described by a
Gaussian distribution function \cite{hobos,boyalamo,hoboydavey} with
\be \langle \xi_{\alpha,\vec k}(t) \rangle =0 ~~;~~ \langle
\xi_{\alpha,\vec k}(t)\xi_{\beta,-\vec k}(t') \rangle =
\mathcal{K}_{\alpha,\beta}(k;t-t') \equiv
 \int_{-\infty}^{\infty}\frac{d\omega}{2\pi} e^{i\omega(t-t')}
\widetilde{\mathcal{K}}_{\alpha \beta}(k;\omega) \label{noisecor}\ee
and the angular brackets denote the averages with the Gaussian
probability distribution function, determined by the averages over
the bath degrees of freedom. The retarded self-energy kernel has the
following spectral representation\cite{hobos} \be \Sigma_{\alpha
\beta}(k;t-t') = \frac{i}{\pi}\int_{-\infty}^{\infty}
e^{i\omega(t-t')}\mathrm{Im}\widetilde{\Sigma}_{\alpha
\beta}(k;\omega) d\omega \label{SEk}\ee where the imaginary part in
the flavor basis is \be \mathrm{Im}\widetilde{\Sigma}(k;\omega) =
\mathrm{Im}\widetilde{\Sigma}_{aa}(k;\omega)\left(
                                             \begin{array}{cc}
                                               1 & 0 \\
                                               0 & 0 \\
                                             \end{array}
                                           \right)\label{Imsig}\,, \ee  and
$\mathrm{Im}\widetilde{\Sigma}_{aa}(k;\omega)$ is obtained from the
cut discontinuity in the one-loop diagram in figure
(\ref{fig:loop}). In this figure the $W$ propagator should be
identified with the \emph{full} charged vector boson propagator in
the standard model, including a   radiative self-energy correction
from a quark, lepton or hadron loop.

Because the bath fields are in thermal equilibrium, the noise
correlation kernel $\widetilde{\mathcal{K}}_{\alpha
\beta}(k;\omega)$ in eqn. (\ref{noisecor}) and the absorptive part
of the retarded self energy $\mathrm{Im}\widetilde{\Sigma}_{\alpha
\beta}(k;\omega)$ obey the generalized fluctuation dissipation
relation\cite{hobos,boyalamo,hoboydavey} \begin{equation}
\widetilde{\mathcal{K}}_{\alpha
\beta}(k;\omega)=\mathrm{Im}\widetilde{\Sigma}_{\alpha
\beta}(k;\omega)\coth\left[\frac{\beta
\omega}{2}\right]\label{flucdiss}
\end{equation}

The solution of the Langevin equation (\ref{langevin})
is\cite{hobos,boyalamo,hoboydavey} \be \Psi_{\alpha,\vk}(t) =
\dot{G}_{\alpha \beta}(k;t)\,\Psi^0_{\beta,\vec k} + {G}_{\alpha
\beta}(k;t)\,
 \Pi^0_{\beta,\vec k} + \int^t_0
G_{\alpha \beta}(k;t')~\xi_{\beta,\vec k}(t-t') dt' \,,
\label{inhosolution}
\end{equation}  from which is clear that the propagator
$G_{\alpha\beta}$ contains all the relevant information for the
non-equilibrium dynamics.

In the Breit-Wigner (narrow width) approximation, the matrix
propagator $G (k;t)$ in the flavor basis is given by\cite{hobos}
\bea G(k;t) = && Z^{(1)}_k \,e^{-\frac{\Gamma_1(k)}{2}t}
\,\Bigg[\frac{\sin(\Omega_1(k)t)}{
{\Omega_1(k)}}\,\mathds{R}^{(1)}(\theta_m )-
\frac{\widetilde{\gamma}(k)}{2}\,\frac{\cos(\Omega_1(k)t)}{\Omega_1(k)}\,\mathds{R}^{(3)}(\theta_m
)\Bigg]+ \nonumber \\ &&Z^{(2)}_k\,
e^{-\frac{\Gamma_2(k)}{2}t}\,\Bigg[
\frac{\sin(\Omega_2(k)t)}{\Omega_2(k)}\,\mathds{R}^{(2)}(\theta_m )+
\frac{\widetilde{\gamma}(k)}{2}\,\frac{\cos(\Omega_2(k)t)}{\Omega_2(k)}\,\mathds{R}^{(3)}(\theta_m
)\Bigg]
                                                       \label{Gdecay}
                                                       \eea where $Z^{(i)}_k$ are the residues at the quasiparticle
                                                       poles and  we have introduced
                the matrices  \be    \mathds{R}^{(1)}
(\theta)   = \Bigg(\begin{array}{cc}
                                                         \cos^2\theta & -\cos\theta \sin \theta \\
                                                         -\cos\theta \sin \theta & \sin^2\theta \\
                                                       \end{array}\Bigg)
                                                       = U(\theta)\, \Bigg(\begin{array}{cc}
                                                         1 & 0 \\
                                                         0 & 0 \\
                                                       \end{array}\Bigg)\,  U^{-1}(\theta)  \label{R1} \ee
                                                       \be  \mathds{R}^{(2)}
                                                       (\theta)    =    \Bigg(\begin{array}{cc}
                                                         \sin^2\theta &  \cos\theta \sin \theta \\
                                                          \cos\theta \sin \theta & \cos^2\theta \\
                                                       \end{array}\Bigg) = U(\theta)\, \Bigg(\begin{array}{cc}
                                                         0 & 0 \\
                                                         0 & 1 \\
                                                       \end{array}\Bigg)\,  U^{-1}(\theta)
                                                       \label{R2}\ee
 \be \label{R3} \mathds{R}^{(3)}(\theta ) = \sin 2\theta ~~ \left(
                                             \begin{array}{cc}
                                               \sin 2\theta  & \cos 2\theta  \\
                                               \cos 2\theta  & -\sin 2\theta  \\
                                             \end{array}
                                           \right) = \sin 2\theta \, U(\theta )~\left(
                                             \begin{array}{cc}
                                               0 & 1 \\
                                               1 & 0 \\
                                             \end{array}
                                           \right)~U^{-1}(\theta )\,. \ee
From the results of reference\cite{hobos} to leading order in $G$,
the mixing angle in the medium is determined from the relations  \be
\cos2\theta_m  = \frac{\cos2\theta -
 \frac{V_{aa}}{\delta M^2}}{  \varrho  } ~~;~~ \sin2\theta_m  =
\frac{\sin2\theta}{ \varrho } \,,\label{tetamedav} \ee where \be
\varrho  = \Bigg[\big(\cos 2\theta -\frac{V_{aa}}{\delta
M^2}\big)^2+\big(\sin2\theta\big)^2\Bigg]^{\frac{1}{2}}\,.
\label{varrho}\ee

The expressions (\ref{tetamedav}) for the mixing angles in the
medium in terms of the mixing angle in the vacuum and the matter
potential is exactly of the same form as in the case of (fermionic)
neutrinos in a medium\cite{book1,book2,book3,raffelt}. An MSW
resonance occurs whenever\cite{book1,book2,book3,raffelt}

\be V_{aa}= \delta M^2 \cos2\theta \,. \label{reso}\ee

The propagating frequencies and widths are given by\cite{hobos} \bea
\Omega_1(k) & = & \omega_1(k)+ \Delta \omega_1(k)
  ~~;~~ \Gamma_1(k) =
\frac{\mathrm{Im}\widetilde{\Sigma}_{aa}(k;\omega_1(k))}{\omega_1(k)}~\cos^2\tm
  \label{omegone}\\
\Omega_2(k) & = &  \omega_2(k)+ \Delta \omega_2(k) ~~;~~\Gamma_2(k)
=
\frac{\mathrm{Im}\widetilde{\Sigma}_{aa}(k;\omega_2(k))}{\omega_2(k)}~\sin^2\tm
\,,\label{omegtwo}\eea  where \bea \omega^2_1(k) & = &
k^2+\overline{M}^2 +\frac{V_{aa}}{2} - \frac{\delta M^2 \, \varrho
}{2} \label{ome1} \\ \omega^2_2(k) & = & k^2+\overline{M}^2
+\frac{V_{aa}}{2} + \frac{\delta M^2 \, \varrho }{2}
\label{ome2}\eea are the propagating frequencies (squared) in the
medium including the matter potential at order $G$, namely the index
of refraction arising from forward scattering, with
$\overline{M}^2\,;\delta M^2$   defined in equation
(\ref{MbarDelM}). The second order frequency shifts are  \bea \Delta
\omega_1(k) & = & -\frac{\cos^2\tm}{2\pi\omega_1(k)} \int d\omega
\mathcal{P}\left[
\frac{\mathrm{Im}\widetilde{\Sigma}_{aa}(k;\omega)}{\omega-\omega_1(k)}\right]\,
\label{delo1}\\\Delta \omega_2(k) & = & -\frac{\sin^2\tm}{2\pi
\omega_2(k)} \int d\omega \mathcal{P}\left[
\frac{\mathrm{Im}\widetilde{\Sigma}_{aa}(k;\omega)}{\omega-\omega_2(k)}\right]\,
\,,  \label{delo2}\eea and\cite{hobos} \be
\label{gamatil}\widetilde{\gamma}(k)=
   \frac{\mathrm{Im}\widetilde{\Sigma}_{aa}(k;\overline{\omega}(k))  }{\omega^2_2(k)-\omega^2_1(k)}~~;~~\overline{\omega}(k)=
    \sqrt{k^2+\overline{M}^2}\,.  \ee The relationship between the
    damping rates $\Gamma_{1,2}$ and the imaginary part of the self
    energy is the same as that obtained in the study of neutrinos
    with standard model interactions in a medium in\cite{hozeno}.

To leading order in  perturbation theory  the denominator in
equation (\ref{gamatil})  is $\delta M^2 \rho $. When the matter
potential dominates (at high temperature in the standard model),
$V_{aa}\gg \delta M^2$ and $\delta M^2 \rho \sim
 V_{aa} \propto G \gg \mathrm{Im}\widetilde{\Sigma}_{aa}\propto G^2$,
thus in this regime $\widetilde{\gamma} \propto G \ll 1$. For
example with active neutrinos with standard model interactions at
high temperature, it was argued in ref.\cite{hozeno} that $ V_{aa}
\propto G_F k T^5/M^2_W$ whereas $\mathrm{Im}\widetilde{\Sigma}_{aa}
\sim G^2_F k T^5$ therefore at high temperature
$\widetilde{\gamma}\sim \mathrm{Im}\widetilde{\Sigma}_{aa}/V_{aa}
\sim g_w \ll 1 $ with $g_w$ the standard model weak coupling.

 In the
opposite limit, for $\delta M^2 \gg V_{aa} \propto G$  the vacuum
mass difference dominates $\rho \sim 1$ and $\widetilde{\gamma} \ll
1$ since $\delta M^2 \gg G \gg G^2$. This analysis is similar to
that in ref.\cite{hozeno} and precludes the possibility of ``quantum
zeno suppression''\cite{stodolsky,kev1} at high temperature.

The \emph{only} region in which $\widetilde{\gamma} $ \emph{may not
be perturbatively small} is near a resonance at which $\rho = |\sin
2\theta|$ and only for very small vacuum mixing angle so that
$\delta M^2 |\sin 2\theta| \propto G^2$. This situation requires a
careful re-examination of the perturbative expansion, and in this
case the propagator \emph{cannot} be described as two separate
Breit-Wigner resonances because the width of the resonances is of
the same order of or larger than the separation between them. Such a
possibility would require a complete re-assessment of the dynamics
of the propagating modes in the medium as a consequence of the
breakdown of the Breit-Wigner (or narrow width) approximation.
However, for very small vacuum mixing angle, indeed a distinct
possibility for $\mathrm{keV}$ sterile neutrinos\cite{kev1}, the MSW
resonance is very narrow and in most of the parameter range
$\widetilde{\gamma} \ll 1$ and can be safely neglected. This is
certainly the case at very high or very low temperature  regimes in
which $V_{aa} \gg \delta M^2$ or $V_{aa}\ll \delta M^2$
respectively.

In summary, it follows from this discussion that
$\widetilde{\gamma}(k) \ll 1$, with the possible exception near an
MSW resonance for extremely small vacuum mixing angle\cite{hobos},
and  such a case must be studied   in detail non-perturbatively.

Hence, neglecting perturbatively small corrections,  the Green's
function in the flavor basis can be written as

\be G(k;t) = U(\tmk) G_m(k;t) U^{-1}(\tmk) \label{trafogreen}\ee
with

\be G_m(k;t) = \Bigg(\begin{array}{cc}
                                                                       e^{-\frac{\Gamma_1(k)}{2}t}
~\frac{\sin(\Omega_1(k)t)}{
{\Omega_1(k)}} & 0 \\
                                                                     0 &   e^{-\frac{\Gamma_2(k)}{2}t}
~ \frac{\sin(\Omega_2(k)t)}{ {\Omega_2(k)}}
                                                                   \end{array}
                                                                   \Bigg)
\label{Gm}\ee

This Green's function and the expression for the damping rates
$\Gamma_{1,2}$ in eqn. (\ref{omegone},\ref{omegtwo}) lead to the
following physical interpretation. The fields that diagonalize the
Green's function on the mass shell, namely $\phi_{1,2}$ are
associated with the quasiparticle modes in the medium and describe
the propagating excitations in the medium. From eqn.
(\ref{trafogreen}) these are related to the flavor fields
$\phi_{a,s}$ by the unitary transformation  \be \phi_a = \cos\tm
\phi_1 + \sin \tm \phi_2 ~~;~~ \phi_s = \cos \tm \phi_2 -\sin\tm
\phi_1\,.\label{shellfields} \ee When the matter potential
$V_{aa}\gg \delta M^2$, namely when $\tm \sim \pi/2$ it follows that
$\phi_a \sim \phi_2$ and the damping rate of the active species is
$\Gamma_2 \sim \Gamma_{aa}$ while $\phi_s \sim \phi_1$ and the
damping rate of the ``sterile'' species is $\Gamma_1 \sim
\Gamma_{aa}\cos^2\tm \ll \Gamma_{aa}$, where \be \Gamma_{aa} \simeq
\frac{\mathrm{Im}\widetilde{\Sigma}_{aa}(k;k)}{k }
\label{gammaaa}\ee is the ultrarelativistic limit of the damping
rate of the active species in absence of mixing. In the opposite
limit, when the medium mixing angle is small $\tm \sim 0$,
corresponding to the near-vacuum case, $\phi_a\sim \phi_1$ and the
active species has a damping rate $\Gamma_1 \sim \Gamma_{aa}$ while
$\phi_s \sim \phi_2$ with $\Gamma_2 \sim \Gamma_{aa} \sin^2 \tm \ll
\Gamma_{aa}$. In both limits the sterile species is weakly coupled
to the plasma, active and sterile species become equally coupled
near an MSW resonance for $\tm \sim \pi/4$.

We emphasize that the relation (\ref{shellfields}) \emph{is not} a
relation between wave functions, but between  the \emph{fields}
associated with the flavor eigenstates (active-sterile) and those
associated with the propagating (quasiparticle) excitations in the
medium (see section (\ref{sec:transprob})).


\section{Quantum kinetics:}\label{sec:QK}

The distribution functions for the active $(a)$ and sterile $(s)$
species are  defined in terms of the diagonal entries of  the mass
matrix in the flavor representation, namely \be N_{\alpha}(k;t) =
\frac{1}{2W_{\alpha}(k)}\Bigg[ \langle
\dot{\phi}_{\alpha}(\vk;t)\dot{\phi}_{\alpha}(-\vk,t)\rangle +
W^2_{\alpha}(k)\langle
 {\phi}_{\alpha}(\vk;t) {\phi}_{\alpha}(-\vk,t)\rangle
 \Bigg]-\frac{1}{2} ~~;~~\alpha=a,s \label{numbers}\ee where \be
 W^2_{\alpha}(k) = k^2+ M^2_{\alpha\alpha} \,.\label{barefreqs} \ee
 The equal time expectation values of Heisenberg field operators are in the initial density
 matrix, and as shown in references\cite{hobos,boyalamo,hoboydavey}
 they are the same as the equal time expectation value of the center
 of mass Wigner variables $\Psi$, where the expectation value is now
 in terms of the initial density matrix for the system and the
 distribution function of the noise which is determined by the
 thermal bath\cite{hobos,boyalamo,hoboydavey}.
 Therefore the distribution functions for the active
 and sterile species are given by

\be N_{\alpha}(k;t) = \frac{1}{2W_{\alpha}(k)}\Bigg[ \langle
\dot{\Psi}_{\alpha}(\vk;t)\dot{\Psi}_{\alpha}(-\vk,t)\rangle +
W^2_{\alpha}(k)\langle
 {\Psi}_{\alpha}(\vk;t) {\Psi}_{\alpha}(-\vk,t)\rangle
 \Bigg]-\frac{1}{2} ~~;~~\alpha=a,s \label{numbers}\ee   and the
 averages are taken over the initial density matrix of the system and the noise
 probability distribution. This expression combined with eqn.(\ref{inhosolution}) makes
manifest that the full time evolution of the distribution function
is completely determined by the \emph{propagator} $ G_{\alpha
\beta}(k,t)$ obtained from the solution of the effective equations
of motion in the medium\cite{hobos}.

 It proves convenient to introduce a \emph{matrix} of distribution
 functions in terms of a parameter $\Omega$ as follows

\be \mathds{N}_{\alpha\beta}(k,t;\Omega) \equiv
\frac{1}{2\Omega}\Bigg[ \langle
\dot{\Psi}_{\alpha}(\vk;t)\dot{\Psi}_{\beta}(-\vk,t)\rangle +
\Omega^2\langle
 {\Psi}_{\alpha}(\vk;t) {\Psi}_{\beta}(-\vk,t)\rangle
 \Bigg]-\frac{1}{2}\,\delta_{\alpha\beta} \label{Nmatrix} \ee from
 which we extract the active and sterile distribution functions from
 the diagonal elements, namely \be
 N_{a}(k;t)=\mathds{N}_{a,a}(k,t;W_a(k))~;~
 N_{s}(k;t)=\mathds{N}_{s,s}(k,t;W_s(k)) \label{distfun}\ee and the
 off-diagonal elements determine off-diagonal correlation functions of the fields and their
 canonical momenta  in the flavor basis.

 We   consider first the initial density matrix for the system
 $\widehat{\rho}_\Phi(0)$ to be \emph{diagonal in the flavor basis}
 with free field correlations \bea &&  \mathrm{Tr}~\widehat{\rho}_\Phi(0) \Psi^0_{\alpha,\vk}
 \Psi^0_{\beta,-\vk}   =
 \frac{1}{ 2W_\alpha(k)}\,\left[1+2N_\alpha(k;0)\right]
 \delta_{\alpha\beta}
 \label{psiini}\\ &&  \mathrm{Tr}~\widehat{\rho}_\Phi(0) \Pi^0_{\alpha,\vk}
 \Pi^0_{\beta,-\vk}   =
 \frac{W_\alpha(k)}{2 }\,\left[1+2N_\alpha(k;0)\right]
 \delta_{\alpha\beta}\label{pini}\\&&   \mathrm{Tr}~\widehat{\rho}_\Phi(0) \Psi^0_{\alpha,\vk}
 \Pi^0_{\beta,-\vk}   = 0 \label{pipsini}\eea with
 $N_\alpha(k;0)$ being the initial distribution functions for the
 active and sterile species. Different initial conditions will be
 studied below.

 Following the steps described in   appendix (\ref{app:num}) it is convenient to
 write $\mathds{N}(k,t;\Omega) = \mathds{N}^{(I)}(k,t;\Omega)+\mathds{N}^{(\xi)}(k,t;\Omega)$ where
 $\mathds{N}^{(I)}$ depends on the initial conditions but not on the noise $\xi$
 and $\mathds{N}^{(\xi)}$ depends on the noise $\xi$ but not on the initial
 conditions. We find

 \bea \label{NIni} \mathds{N}^{(I)}(k,t;\Omega) & = & \mathds{R}^{(1)}(\tm)\,
 e^{-\Gamma_1 t} \Bigg\{ \cos^2(\tm)
 \left[\frac{W^2_a+\Omega^2_1}{2W_a\Omega}
 \right]\left[\frac{1}{2}+N_a(0)\right]+\sin^2(\tm)
 \left[\frac{W^2_s+\Omega^2_1}{2\Omega W_s}
 \right]\left[\frac{1}{2}+N_s(0)\right]\Bigg\}
 \left[ \frac{\Omega^2+\Omega^2_1}{2\Omega^2_1}\right]   \nonumber \\
 & + & \mathds{R}^{(2)}(\tm)\,e^{-\Gamma_2 t}  \Bigg\{ \sin^2(\tm)
 \left[\frac{W^2_a+\Omega^2_2}{2W_a\Omega}
 \right]\left[\frac{1}{2}+N_a(0)\right]+\cos^2(\tm)
 \left[\frac{W^2_s+\Omega^2_2}{2\Omega W_s}
 \right]\left[\frac{1}{2}+N_s(0)\right]\Bigg\}
 \left[ \frac{\Omega^2+\Omega^2_2}{2\Omega^2_2}\right] \nonumber \\& + &
 \mathds{R}^{(3)}(\tm)\,e^{-\frac{1}{2}(\Gamma_1+\Gamma_2)t}\cos\left[(\Omega_1-\Omega_2)t\right]
 \left[\frac{\Omega^2+\Omega_2\Omega_1}{4\Omega_1\Omega_2}
 \right]\Bigg\{\frac{(W_a-W_s)}{4\Omega}\left(\frac{\Omega_1\Omega_2}{W_aW_s}-1\right)\nonumber \\
 &&+  N_a(0)\left(\frac{\Omega_1\Omega_2+W^2_a}{2\Omega W_a}  \right)-  N_s(0)
 \left(\frac{\Omega_1\Omega_2+W^2_s}{2\Omega W_s}  \right)
 \Bigg\}-\frac{\mathds{1}}{2}\,.\eea  We have suppressed the dependence on $k$
 to simplify the notation.  The contribution from the noise term can be written as \be
\mathds{N}^{(\xi)}(k,t;\Omega) = \frac{1}{2\Omega}
\int\frac{d\omega}{2\pi} U(\tm)\Bigg\{h_m(\omega,t)
\mathcal{K}_m(\omega) h^*_m(\omega,t) + \Omega^2 f_m(\omega,t)
\mathcal{K}_m(\omega) f^*_m(\omega,t)\Bigg\}
U^{-1}(\tm)\label{Nnoise}\ee where

\be h_m(\omega,t) = \int_0^t e^{-i\omega t'} G_m(k;t') ~~;~~
f_m(\omega,t) = \int_0^t e^{-i\omega t'}
\dot{G}_m(k;t')\label{hf}\ee and

\be\mathcal{K}_m(\omega) =
\mathrm{Im}\widetilde{\Sigma}_{aa}(k;\omega)\,\left[1+2n(\omega)
\right] \Bigg(\begin{array}{cc}
                                                            \cos^2(\tm) & \cos(\tm)\sin(\tm) \\
                                                            \cos(\tm)\sin(\tm) &
                                                            \sin^2(\tm)
                                                          \end{array}
 \Bigg)
\label{Km} \ee After lengthy but straightforward algebra we find
\bea \mathds{N}^{(\xi)}(k,t;\Omega) =  &&  \left[
\frac{\Omega^2+\Omega^2_1}{2\Omega_1\Omega} \right]\left[
\frac{1}{2}+ n(\Omega_1(k))\right]\left(1-e^{-\Gamma_1(k)t}\right)\,
\mathds{R}^{(1)}(\tmk) +\nonumber \\&&  \left[
\frac{\Omega^2+\Omega^2_2}{2\Omega_2\Omega} \right]\left[
\frac{1}{2}+ n(\Omega_2(k))\right]\left(1-e^{-\Gamma_2(k)t}\right)\,
\mathds{R}^{(2)}(\tmk) \label{Nchi}\eea where we have neglected
terms proportional to $\widetilde{\gamma}$.

{\bf Approximations:} In arriving at the expressions (\ref{NIni}),
(\ref{Nchi}), we have made the following approximations:
\begin{itemize} \item{{\bf (a)} We have taken $Z^{(i)}_k =1$ thus neglecting
terms which are perturbatively small, of $\mathcal{O}(G^2)$.}
\item{{\bf (b)} We have  assumed  $\Gamma_i/\Omega_i \ll1$, which is warranted in perturbation
theory and neglected terms proportional to this ratio. }\item{ {\bf
(c)} As discussed above, consistently with perturbation theory we
have assumed $\widetilde{\gamma}(k) \ll1$ and neglected terms
proportional to it. This corresponds to the interaction rate much
smaller than the oscillation frequencies and relies on the
consistency of the perturbative expansion. }\item{{\bf (d)} In
oscillatory terms we have taken a time average over the \emph{rapid
time scales} $1/\Omega_{1,2}$ replacing $\sin^2( \Omega_{1,2}t) =
\cos^2 (
 \Omega_{1,2}t) \rightarrow  1/2\,;\,\sin ( \Omega_{1,2}t) = \cos  (
 \Omega_{1,2}t)\rightarrow 0$.}
 \end{itemize}

 {\bf Ultrarelativistic limit:}
The above expressions simplify considerably in the
\emph{ultrarelativistic limit} in which \be \Omega \sim W_a(k) \sim
W_s(k) \sim \Omega_1(k) \sim \Omega_2(k) \sim k \,,\label{URlim}\ee
and in this limit it follows that \be \Gamma_1 =
\Gamma_{aa}\cos^2\tm~;~\Gamma_2=\Gamma_{aa}\sin^2\tm
\label{URgamas}\ee and  $\Gamma_{aa}$ is the ultrarelativistic limit
of the width of the active species in the absence of mixing given by
eqn. (\ref{gammaaa}). In this limit we obtain the following simple
expression for the time evolution of the occupation number
\emph{matrix} in the flavor basis (suppressing the $k$ dependence
for simplicity)

\bea \mathds{N}(t) & = &
\mathds{R}^{(1)}(\tm)\Bigg[n(\Omega_1)+\Big(N_a(0)\cos^2(\tm)+N_s(0)\sin^2(\tm)-n(\Omega_1)\Big)\,e^{-\Gamma_1
t}\Bigg] \nonumber \\ & + &
\mathds{R}^{(2)}(\tm)\Bigg[n(\Omega_2)+\Big(N_a(0)\sin^2(\tm)+N_s(0)\cos^2(\tm)-n(\Omega_2)\Big)\,e^{-\Gamma_2
t}\Bigg] \nonumber \\ &+& \frac{1}{2}~\mathds{R}^{(3)}(\tm)
\,e^{-\frac{1}{2}(\Gamma_1+\Gamma_2)t}\,\cos\left[(\Omega_1-\Omega_2)t\right]
\Big(N_a(0)-N_s(0) \Big) \,.\label{numbermtx}\eea It is
straightforward to verify that \be \mathds{N}(0) = \left(
                                                     \begin{array}{cc}
                                                       N_a(0) & 0 \\
                                                       0 & N_s(0) \\
                                                     \end{array}
                                                   \right)\,. \ee

The active and sterile populations are given by the diagonal
elements of  (\ref{numbermtx}), namely \bea N_a(t) & = &
\cos^2(\tm)\Bigg[n(\Omega_1)+\Big(N_a(0)\cos^2(\tm)+N_s(0)\sin^2(\tm)-n(\Omega_1)\Big)\,e^{-\Gamma_1
t}\Bigg] \nonumber \\ & + &
\sin^2(\tm)\Bigg[n(\Omega_2)+\Big(N_a(0)\sin^2(\tm)+N_s(0)\cos^2(\tm)-n(\Omega_2)\Big)\,e^{-\Gamma_2
t}\Bigg] \nonumber \\ &+& \frac{1}{2}~\sin^2(2\tm)
\,e^{-\frac{1}{2}(\Gamma_1+\Gamma_2)t}\,\cos\left[(\Omega_1-\Omega_2)t\right]
\Big(N_a(0)-N_s(0) \Big) \,.\label{numberact}\eea

\bea N_s(t) & = &
\sin^2(\tm)\Bigg[n(\Omega_1)+\Big(N_a(0)\cos^2(\tm)+N_s(0)\sin^2(\tm)-n(\Omega_1)\Big)\,e^{-\Gamma_1
t}\Bigg] \nonumber \\ & + &
\cos^2(\tm)\Bigg[n(\Omega_2)+\Big(N_a(0)\sin^2(\tm)+N_s(0)\cos^2(\tm)-n(\Omega_2)\Big)\,e^{-\Gamma_2
t}\Bigg] \nonumber \\ &-& \frac{1}{2}~\sin^2(2\tm)
\,e^{-\frac{1}{2}(\Gamma_1+\Gamma_2)t}\,\cos\left[(\Omega_1-\Omega_2)t\right]
\left(N_a(0)-N_s(0) \right) \,.\label{numberste}\eea

 The oscillatory term which results from the
  interference of the propagating modes $1,2$ damps out with a damping factor \be
\frac{1}{2}(\Gamma_1+\Gamma_2) = \frac{\Gamma_{aa}}{2} \label{deco}
\ee which determines the \emph{decoherence time scale} $\tau_{dec} =
2/\Gamma_{aa}$. These expressions are one of the main results of
this article.

{\bf Initial density matrix diagonal in the $1-2$ basis:} The above
results were obtained assuming that the initial density matrix is
diagonal in the flavor basis, if instead, it is diagonal in the
basis of the propagating modes in the medium, namely the $1-2$
basis, it is straightforward to find the result \bea \mathds{N}(t)
= && U(\tm)\Bigg\{ \left(
                                     \begin{array}{cc}
                                       1 & 0 \\
                                       0 & 0 \\
                                     \end{array}
                                   \right)
  \Bigg[n(\Omega_1)+\Big(N_1(0) -n(\Omega_1)\Big)\,e^{-\Gamma_1
t}\Bigg] \nonumber \\ && + \left( \begin{array}{cc}
                                       0 & 0 \\
                                       0 & 1 \\
                                     \end{array}\right)
 \Bigg[n(\Omega_2)+\Big(N_2(0) -n(\Omega_2)\Big)\,e^{-\Gamma_2
t}\Bigg] \Bigg \}U^{-1}(\tm)\,.\label{numbermtx12}\eea

In particular, the active and sterile distribution functions become
\bea N_a(t) & = &  \cos^2\tm \Bigg[n(\Omega_1)+\Big(N_1(0)
-n(\Omega_1)\Big)\,e^{-\Gamma_1 t}\Bigg]+\sin^2\tm
\Bigg[n(\Omega_2)+\Big(N_2(0) -n(\Omega_2)\Big)\,e^{-\Gamma_2
t}\Bigg]\label{Nadiag}\\N_s(t) & = &
\cos^2\tm\Bigg[n(\Omega_2)+\Big(N_2(0)
-n(\Omega_2)\Big)\,e^{-\Gamma_2 t}\Bigg] + \sin^2\tm
\Bigg[n(\Omega_1)+\Big(N_1(0) -n(\Omega_1)\Big)\,e^{-\Gamma_1
t}\Bigg]\,.\label{Nsdiag}\eea

The   results summarized by eqns. (\ref{numberact}-\ref{Nsdiag})
show that the distribution functions for the propagating modes in
the medium, namely the $1,2$ quasiparticles, reach equilibrium with
the damping factor $\Gamma_{1,2}$ which is \emph{twice} the damping
rate of the quasiparticle modes (see eqn. (\ref{Gm})).  The
interference term is present \emph{only} when the initial density
matrix is \emph{off diagonal} in the (1,2) basis of propagating
modes in the medium.

If the initial density matrix is off-diagonal in the (1,2) basis,
these off diagonal components damp-out within the decoherence time
scale $\tau_{dec}$, while the diagonal elements attain the values of
the equilibrium distributions on the time scales
$1/\Gamma_1,1/\Gamma_2$.

\section{The Quantum Master equation}\label{sec:QME}
The quantum master equation is the equation of motion of the
\emph{reduced} density matrix of the system fields in the
interaction picture after integrating out the bath degrees of
freedom. The first step is to define the interaction picture, for
which   a precise separation between the free and interaction parts
in the Hamiltonian is needed\cite{qobooks}. In order to carry out
the perturbative expansion in terms of the eigenstates in the
medium, we include the lowest order forward scattering correction,
namely the index of refraction into the un-perturbed Hamiltonian.
This is achieved by writing the term \be \phi^2_a \chi^2 = \phi^2_a
\langle \chi^2 \rangle + \phi^2_a \delta \chi^2 \ee where \be \delta
\chi^2 = \chi^2-\langle \chi^2 \rangle~~;~~\langle \delta \chi^2
\rangle =0 \ee and the average is performed in the bath density
matrix $\rho_B(0) = e^{-\beta H_0[\chi,W]}$. In this manner the
quadratic part of the Lagrangian density for the active and sterile
fields is \be \mathcal{L}_0[\phi] = \frac{1}{2} \left\{
\partial_{\mu} \Phi^T
\partial^{\mu} \Phi -   \Phi^T  (\mathds{M}^2 + \mathds{V})   \Phi \right\}
\ee where $\mathds{V}$ is the  matter potential   given by eqn.
(\ref{matV}). The unperturbed Hamiltonian for the system fields in
the medium is diagonalized by the unitary transformation
(\ref{trafo}) but with the unitary matrix $U(\tm)$ with $\tm$ being
the mixing angle in the medium given by equations
(\ref{tetamedav},\ref{varrho}) and $\varphi_1,\varphi_2$ are now the
fields associated with the eigenstates of the Hamiltonian in the
medium including the index of refraction correction from the matter
potential to $\mathcal{O}(G)$ ($\mathcal{O}(G_F)$ in the case of
neutrinos with standard model interactions). Introducing creation
and annihilation operators for the fields $\varphi_{1,2}$ with usual
canonical commutation relations, the unperturbed Hamiltonian
\emph{for the propagating modes in the medium  including the index
of refraction }is \be H_S[\varphi_{1,2}] =
\sum_{\vk}\sum_{i=1,2}\left[a^\dagger_i(\vk)a_i(\vk)\omega_i(k)\right]\label{HoS}\ee
where $\omega_i(k)$ are the propagating frequencies in the medium
given in equation (\ref{ome1},\ref{ome2}). The interaction
Hamiltonian is \be H_I = G\int d^3x\left[ \phi^2_a \delta \chi^2 +
\phi_a \mathcal{O}_a\right] \label{HI}\ee where \be \phi_a =
\cos(\tm) \varphi_1+\sin(\tm) \varphi_2\,.\label{fia}\ee This
formulation represents a \emph{re-arrangement} of the perturbative
expansion in terms of the fields that create and annihilate the
propagating modes \emph{in the medium}. The remaining steps are
available in the quantum optics literature\cite{qobooks}. Denoting
the Hamiltonian for the bath degrees of freedom $H_0[\chi,W] \equiv
H_B$ the total Hamiltonian is $H=H_S+H_B+H_I \equiv
\overline{H}_0+H_I$. The density matrix in the interaction picture
is \be \widehat{\rho}_{i}(t) = e^{i\overline{H}_0 t}
\widehat{\rho}(t)e^{-i\overline{H}_0 t}\label{rhoip}\ee where
$\widehat{\rho}(t)$ is given by eqn. (\ref{rhooft}) and it obeys the
equation of motion \be \frac{d\widehat{\rho}_{i}(t)}{dt} =
-i\left[H_{I}(t),\widehat{\rho}_{i}(t)\right] \label{eqrhoip}\ee
with $H_I(t) =e^{i\overline{H}_0 t} H_I e^{-i\overline{H}_0 t}$ is
the interaction Hamiltonian in the interaction picture of
$\overline{H}_0$. Iteration of this equation up to second order in
the interaction yields\cite{qobooks} \be
\frac{d\widehat{\rho}_{i}(t)}{dt} =
-i\left[H_{I}(t),\widehat{\rho}_{i}(0)\right]- \int^t_0 dt'\left[
 H_I(t),\left[H_I(t'),\widehat{\rho}_{i}(t')\right]\right]+\cdots
\label{2ndord}\ee

The \emph{reduced} density matrix for the system is obtained from
the total density matrix by tracing over the bath degrees of freedom
which are assumed to remain in equilibrium\cite{qobooks}. At this
stage, several standard approximations are invoked\cite{qobooks}:
\begin{itemize} \item   \textbf{i): factorization:}  the total density matrix is assumed
to factorize \be\widehat{\rho}_i(t)=
\rho_{S,i}(t)\otimes\rho_B(0)\label{fact}\ee where it is assumed
that the bath remains in equilibrium, this approximation is
consistent    with  obtaining the effective action by  tracing over
the bath degrees of freedom with an equilibrium thermal density
matrix. The correlation functions of the bath degrees of freedom are
not modified by the coupling to the system.\item \textbf{ii):
Markovian approximation:} the memory of the evolution is neglected
and in the double commutator in (\ref{2ndord})
$\widehat{\rho}_i(t')$ is replaced by $\widehat{\rho}_i(t)$ and
taken out of the integral. \end{itemize}   Taking the trace over the
bath degrees of freedom   yields the quantum master equation for the
reduced density matrix, \be \frac{d {\rho}_{S,i}(t)}{dt} =  -
\int^t_0 dt'{\mathrm{Tr}}_B\big\{\left[
 H_I(t),\left[H_I(t'),\widehat{\rho}_{i}(t)\right]\right]\big\}+\cdots
\label{2ndordred}\ee where the first term has vanished by dint of
the fact that the matter potential was absorbed into the unperturbed
Hamiltonian, namely $Tr_B \rho_B(0) \delta \chi^2 =0$. This is an
important aspect of the interaction picture in the basis of the
propagating states \emph{in the medium}. Up to second order   we
will only consider the interaction term \be H_I(t) = \sum_{\vk}
\left[\cos\tm \varphi_{1,\vk}(t)+\sin\tm
\varphi_{2,\vk}(t)\right]\mathcal{O}_{-\vk}(t) \label{inter}\ee
where we have written the interaction Hamiltonian in terms of
spatial Fourier transforms and the fields are in the interaction
picture of $\overline{H}_0$. We neglect non-linearities from the
second order contributions of the term $\phi^2_a \delta \chi^2$, the
non-linearities associated with the neutrino background are included
in the forward scattering corrections accounted for in the matter
potential. The quartic non-linearities are associated with active
``neutrino-neutrino''  elastic scattering and are not relevant for
the production of the sterile species.

The next steps   are: i) writing out explicitly the nested
commutator in (\ref{2ndordred}) yielding  four different terms, ii)
taking the trace over the bath degrees of freedom yielding   the
correlation functions of the bath operators $Tr_B \mathcal{O}(t)
\mathcal{O}(t')$ (and $t \leftrightarrow t'$) and, iii)   carrying
out the integrals in the variable $t'$. While straightforward these
steps are lengthy and technical and are relegated to   appendix
(\ref{app:qme}). Two further approximations are
invoked\cite{qobooks}, \begin{itemize}   \item \textbf{iii): the
``rotating wave approximation'':} terms that feature rapidly varying
phases of the form $e^{\pm   2 i\omega_{1,2} t};e^{\pm     i
(\omega_1+\omega_2) t}$ are averaged out in time leading to their
cancellation. This approximation  also has a counterpart in the
effective action approach in   the averaging of rapidly varying
terms, see the discussion after equation (\ref{Nchi}). \item
\textbf{ iv): the Wigner Weisskopf approximation:} time integrals of
the form \be \int^t_0 e^{-i(\omega-\Omega)\tau} d\tau \approx
-i\mathcal{P}\left[ \frac{1}{\omega-\Omega}\right] + \pi
\delta(\omega-\Omega)\,\label{WW}\ee where $\mathcal{P}$ stands for
the principal part. The Markovian approximation   \textbf{(ii)} when
combined with the Wigner-Weisskopf approximation  is equivalent to
approximating the propagators by their narrow width Breit-Wigner
form in the effective action. \end{itemize}

 All of these approximations \textbf{i)}- \textbf{iv)} detailed
above are standard in the derivation of quantum master equations in
the literature\cite{qobooks}.

The quantum master equation is obtained  in appendix
(\ref{app:qme}), it features diagonal and off-diagonal terms in the
$1-2$ basis and is of the Lindblad form\cite{qobooks} which ensures
that the trace of the reduced density matrix is a constant of motion
as it must be, because it is consistently derived from the full
Liouville evolution (\ref{rhooft}).  We now focus on the
ultrarelativistic case $\omega_1(k) \sim \omega_2(k) \sim k$ which
leads to substantial simplifications and is the relevant case for
sterile neutrinos in the early Universe, we also neglect the second
order corrections to the propagation frequencies. With these
simplifications we obtain,

  \bea \label{qmefin}  \frac{d {\rho}_{S,i}
}{dt} & = & \Bigg\{ \sum_{j=1,2} \sum_{\vk}  - \frac{\Gamma_j(k)}{2}
\Bigg[\big[1+n(\omega_j(k))\big]\bigg( \rho_{S,i}a^\dagger_j(\vk)
a_j(\vk)+a^\dagger_j(\vk) a_j(\vk)\rho_{S,i}   -  2
a_j(\vk)\rho_{S,i}a^\dagger_j(\vk) \bigg)\nonumber \\ & + &
n(\omega_j(k)) \bigg( \rho_{S,i}a_j(\vk) a^\dagger_j(\vk)+a_j(\vk)
a^\dagger_j(\vk)\rho_{S,i}-2a^\dagger_j(\vk)\rho_{S,i}a_j(\vk)
\bigg) \Bigg] \nonumber \\ & - & \sum_{\vk} \frac{\widetilde{\Gamma}
(k)}{2}\Bigg\{
  \Bigg[\Big(1+n(\omega_1(k))\Big)
\Big(a^\dagger_2(k;t)a_1(k;t)\rho_{S,i}+\rho_{S,i}a^\dagger_1(k;t)a_2(k;t)-a_2(k;t)\rho_{S,i}a^\dagger_1(k;t) \nonumber \\
 & - & a_1(k;t)\rho_{S,i}a^\dagger_2(k;t)\Big)  + n(\omega_1(k)) \Big(a_2(k;t)a^\dagger_1(k;t)\rho_{S,i}
 +\rho_{S,i}a_1(k;t)a^\dagger_2(k;t)-a^\dagger_2(k;t)\rho_{S,i}a_1(k;t) \nonumber \\
 & - & a^\dagger_1(k;t)\rho_{S,i}a_2(k;t)\Big)\Bigg] + \Bigg[\Big(1+n(\omega_2(k))
 \Big)\Big(a^\dagger_1(k;t)a_2(k;t)\rho_{S,i}+\rho_{S,i}a^\dagger_2(k;t)a_1(k;t)-a_1(k;t)\rho_{S,i}a^\dagger_2(k;t)
  \nonumber \\
 & - & a_2(k;t)\rho_{S,i}a^\dagger_1(k;t)\Big)  + n(\omega_2(k)) \Big(a_1(k;t)a^\dagger_2(k;t)\rho_{S,i}
 +\rho_{S,i}a_2(k;t)a^\dagger_1(k;t)-a^\dagger_1(k;t)\rho_{S,i}a_2(k;t) \nonumber \\
 & - & a^\dagger_2(k;t)\rho_{S,i}a_1(k;t)\Big)\Bigg]\Bigg\} \,, \eea
 where \bea \Gamma_1(k) = \Gamma_{aa}(k) \cos^2\tm ~~,~~\Gamma_2(k)
 = \Gamma_{aa}(k)\sin^2\tm \nonumber \\\widetilde{\Gamma} (k) =
 \frac{1}{2}\sin2\tm \Gamma_{aa}(k) ~~;~~ \Gamma_{aa}(k) =\frac{
 \mathrm{Im}\Sigma_{aa}(k,k)}{k} \label{gammas}\eea and the
 interaction picture operators are given in eqn. (\ref{ipas}).
The expectation value of any system's operator $A$ is given by \be
\langle A \rangle(t) = \mathrm{Tr}\rho_{i,S}(t)A(t) \label{aveA}\ee
where $A(t)$ is the operator in the interaction picture of
$\overline{H}_0$, thus the time derivative of this expectation value
contains two contributions \be \frac{d}{dt}\langle A \rangle(t)=
\mathrm{Tr}\dot{\rho}_{i,S}(t)A(t)+\mathrm{Tr}\rho_{i,S}(t)\dot{A}(t)\,
. \label{timeder}\ee The distribution functions for active and
sterile species is defined as in  equation (\ref{numbers}) with the
averages defined as in (\ref{aveA}), namely \be N_{\alpha}(k;t)=
\mathrm{Tr}\rho_{i,S}(t) \Bigg[\frac{
\dot{\phi}_{\alpha}(\vk;t)\dot{\phi}_{\alpha}(-\vk,t)}
{2W_{\alpha}(k)}+ \frac{W_{\alpha}(k)}{2}~
 {\phi}_{\alpha}(\vk;t) {\phi}_{\alpha}(-\vk,t)
 \Bigg]-\frac{1}{2}\, ,\label{numbersII}\ee where the fields are in the interaction picture of $\overline{H}_0$.
 The active and sterile fields are related to the fields that create and annihilate the
propagating modes in the medium as \be \phi_a(\vk) = \cos\tm
\varphi_1(\vk)+ \sin\tm \varphi_2(\vk)~~;~~\phi_s(\vk) = \cos\tm
\varphi_2(\vk)- \sin\tm \varphi_1(\vk)\,. \label{fieldrel}\ee In the
interaction picture of $\overline{H}_0$ \be \varphi_j(\vk,t) =
\frac{1}{\sqrt{2\omega_j(k)}}\Bigg[a_j(\vk)~e^{-i\omega_j(k)t}+a^\dagger_j(-\vk)~e^{
i\omega_j(k)t} \Bigg]\label{ipfields}\ee where $\omega_j(k)$ are the
propagation frequencies in the medium up to leading order in $G$,
given by equations (\ref{ome1},\ref{ome2}). Introducing this
expansion into the expression (\ref{numbersII}) we encounter the
ratio of the propagating frequencies in the medium $\omega_j$ and
the bare frequencies $W_{\alpha}$. Just as we did in the previous
section, we focus on the relevant case of ultrarelativistic species
and approximate as in equation (\ref{URlim}) $\omega_j(k) \sim
W_\alpha(k) \sim k$, in which case we find the relation between the
creation-annihilation operators for the flavor fields and those of
the $1,2$ fields to be\cite{hobos} \be a_{a}(\vk,t) = \cos\tm
a_1(\vk,t)+ \sin \tm a_2(\vk,t)~~;~~a_{s}(\vk,t) = \cos\tm
a_2(\vk,t)- \sin \tm a_1(\vk,t) \label{as}\ee leading to the simpler
expressions for the active and sterile distributions, \bea N_a(k;t)
= \mathrm{Tr}\rho_{i,S}(t)\Bigg[ && \cos^2\tm
a^\dagger_1(\vk,t)a_1(\vk,t)+ \sin^2\tm
a^\dagger_2(\vk,t)a_2(\vk,t)\nonumber \\ &&  + \frac{1}{2}\sin2\tm
(a^\dagger_1(\vk,t)a_2(\vk,t)+a^\dagger_2(\vk,t)a_1(\vk,t))\Bigg]\label{nac}\eea
\bea  N_s(k;t)   =   \mathrm{Tr}\rho_{i,S}(t)\Bigg[&& \cos^2\tm
a^\dagger_2(\vk,t)a_2(\vk,t)+ \sin^2\tm
a^\dagger_1(\vk,t)a_1(\vk,t)\nonumber \\ &   & - \frac{1}{2}\sin2\tm
(a^\dagger_1(\vk,t)a_2(\vk,t)+a^\dagger_2(\vk,t)a_1(\vk,t))\Bigg]\,.\label{nst}\eea
In the interaction picture of $\overline{H}_0$ the products
$a^\dagger_j(\vk,t)a_j(\vk,t)$ are time independent and
$a^\dagger_1(\vk,t)a_2(\vk,t)=
a^\dagger_1(\vk,0)a_2(\vk,0)~e^{i(\omega_1(k)-\omega_2(k))t}$. It is
convenient to introduce the distribution functions and off-diagonal
correlators \bea  &&    n_{11}(k,t) = \mathrm{Tr}\rho_{i,S}(t)
a^\dagger_1(k,t)a_1(k,t)~~,~~n_{22}(k,t) = \mathrm{Tr}\rho_{i,S}(t)
a^\dagger_2(k,t)a_2(k,t) \label{distfun} \\ &&  n_{12}(k,t) =
\mathrm{Tr}\rho_{i,S}(t) a^\dagger_1(k,t)a_2(k,t) ~~,~~n_{ 21}(k,t)
= \mathrm{Tr}\rho_{i,S}(t) a^\dagger_2(k,t)a_1(k,t)= n^*_{1 2}(k,t)
\,. \label{correfun}\eea In terms of these, the distribution
functions for the active and sterile species in the
\emph{ultrarelativistic} limit becomes \bea N_a(k;t)  & =  &
\cos^2\tm n_{11}(k;t)+\sin^2\tm n_{22}(k;t)+ \frac{1}{2}\sin 2\tm
\Big(n_{12}(k;t)+n_{21}(k;t)\Big) \label{nacti}\\N_s(k;t)  & =  &
\sin^2\tm n_{11}(k;t)+\cos^2\tm n_{22}(k;t)- \frac{1}{2}\sin 2\tm
\Big(n_{12}(k;t)+n_{21}(k;t)\Big) \,. \label{stenu}\eea  From eqn.
(\ref{timeder}) we obtain the following kinetic equations for
$n_{ij}(k;t)$ \bea \dot{n}_{11} & = &
-\Gamma_{1}\Big[n_{11}-n_{eq,1}\Big] - \frac{\widetilde{\Gamma}}{2}
\Big[n_{12}+n_{21}\Big] \label{dotn11}
\\ \dot{n}_{22}  & = &  -\Gamma_{2}\Big[n_{22}-n_{eq,2}\Big] -
\frac{\widetilde{\Gamma}}{2} \Big[n_{12}+n_{21}\Big]
\label{dotn22}\\\dot{n}_{12} & = & \Big[-i\Delta\omega
-\frac{\Gamma_{aa}}{2}\Big]n_{12} -
\frac{\widetilde{\Gamma}}{2}\Big[(n_{11}-n_{eq,1}) +
(n_{22}-n_{eq,2}) \Big] \label{dotn12}\\ \dot{n}_{21} & = & \Big[+
i\Delta\omega -\frac{\Gamma_{aa}}{2}\Big]n_{21} -
\frac{\widetilde{\Gamma}}{2}\Big[(n_{11}-n_{eq,1}) +
(n_{22}-n_{eq,2}) \Big] \,,\label{dotn21}\eea where   $ n_{eq,j} =
n(\omega_j(k))$ are the equilibrium distribution functions for the
corresponding propagating modes, and $\Delta \omega =
(\omega_2(k)-\omega_1(k))$. As we have argued above, in perturbation
theory $\Gamma_{aa}(k) /\Delta \omega(k) \ll 1$, which is the same
statement as the approximation $\widetilde{\gamma}\ll 1$ as
discussed for the effective action, and in this case the off
diagonal contributions to the kinetic equations yield perturbative
corrections to the distribution functions and correlators. To
leading order in this ratio we find the distribution functions,

\bea && n_{11}( t)= n_{eq,1} + \Big(n_{11}( 0)-n_{eq,1}
\Big)~e^{-\Gamma_1  t}
-\frac{\widetilde{\Gamma}e^{-\Gamma_1t}}{2\Delta \omega}\Bigg[i~
n_{12}(0)\Big[e^{-i\Delta \omega
t}~e^{\frac{1}{2}(\Gamma_2-\Gamma_1)t}-1\Big]+c.c.\Bigg]
\label{n1oft}\\
&& n_{22}( t)= n_{eq,2} + \Big(n_{22}( 0)-n_{eq,2}
\Big)~e^{-\Gamma_2 t}
-\frac{\widetilde{\Gamma}e^{-\Gamma_2t}}{2\Delta \omega}\Bigg[i~
n_{12}(0)\Big[e^{-i\Delta \omega
t}~e^{\frac{1}{2}(\Gamma_1-\Gamma_2)t}-1\Big]+c.c.\Bigg]\label{n2oft}\eea
and off-diagonal correlators \bea   n_{12}( t) & = &    e^{-i\Delta
\omega t}~e^{-\frac{\Gamma_{aa}}{2}  t}~\Bigg\{n_{12}(0) +i
 \frac{\widetilde{\Gamma}}{2\Delta \omega}\Bigg[ (n_{11}(0)-n_{eq,1})\Big[e^{i\Delta \omega
t}~ e^{ \frac{1}{2}(\Gamma_2-\Gamma_1)t}-1\Big]\nonumber
\\&+&(n_{22}(0)-n_{eq,2})\Big[e^{i\Delta \omega t}~ e^{
\frac{1}{2}(\Gamma_1-\Gamma_2)t}-1\Big]\Bigg]\Bigg\}\label{n12oft}\eea
where \be \frac{\widetilde{\Gamma}}{2\Delta \omega} =
\frac{1}{2}\sin2\tm
\frac{\mathrm{Im}\Sigma_{aa}(k,k)}{2k(\omega_2(k)-\omega_1(k)} =
\frac{1}{2}\sin2\tm \widetilde{\gamma}\ee with $\widetilde{\gamma}$
defined in eqn. (\ref{gamatil}) and we have suppressed the momenta
index for notational convenience.

\subsection{Comparing the effective action and quantum master
equation}

We can now establish the equivalence between the time evolution of
the distribution functions obtained from the effective action and
the quantum master equation, however in order to compare the results
we must first determine the initial conditions in equations
(\ref{n1oft}-\ref{n12oft}). The initial values $n_{ij}(0)$ must be
determined from the initial condition and depend on the initial
density matrix. Two important cases stand out: i) an initial density
matrix diagonal in the flavor basis or ii) diagonal in the $1-2$
basis of propagating eigenstates in the medium.

\vspace{2mm}

{\bf Initial density matrix diagonal in the flavor basis:}   the
initial expectation values are obtained by inverting the relation
between $\varphi_{1,2}$ and $\phi_{a,s}$. We obtain \bea n_{11}(0)=
 \langle a^\dagger_1(\vk )a_1(\vk ) \rangle(0) &  = &  \cos^2\tm
N_a(0)+\sin^2\tm N_s(0) \label{ini1} \\ n_{22}(0)=
 \langle a^\dagger_2(\vk )a_2(\vk ) \rangle(0) &  = &  \cos^2\tm
N_s(0)+\sin^2\tm N_a(0)\label{ini2}\\n_{12}(0)= \langle
a^\dagger_1(\vk )a_2(\vk ) \rangle (0) & = &  \frac{1}{2} \sin 2\tm
(N_a(0)-N_s(0)) \label{ini12}\eea It is straightforward to establish
the equivalence between the results obtained from the effective
action and those obtained above from the quantum master equation as
follows: i) neglect the second order frequency shifts ($\Omega_{1,2}
\sim \omega_{1,2}$) and the perturbatively small corrections of
order $\widetilde{\gamma}$, ii) insert the  initial conditions
(\ref{ini1}-\ref{ini12}) in the solutions
(\ref{n1oft}-\ref{n12oft}), finally  using the relations
(\ref{nacti},\ref{stenu}) for the active and sterile distribution
functions we find precisely the results given by equations
(\ref{numberact},\ref{numberste}) obtained via the non-equilibrium
effective action.

\vspace{2mm}

{\bf Initial density matrix diagonal in the $1-2$ basis:} in this
case \bea \langle a^\dagger_1(\vk )a_1(\vk ) \rangle(0) &  = &
N_1(0) \nonumber \\ \langle a^\dagger_2(\vk )a_2(\vk ) \rangle(0) &
= & N_2(0) \nonumber \\ \langle a^\dagger_1(\vk )a_2(\vk ) \rangle
(0) & = & 0 \label{inidia12}\eea with these initial conditions it is
straightforward to  obtain the result (\ref{Nadiag},\ref{Nsdiag}).

The fundamental advantage in the method of the effective action is
that it highlights that the main ingredient is the \emph{full
propagator in the medium} and the emerging time scales for the time
evolution of distribution functions and coherences are completely
determined by the quasiparticle dispersion relations and damping
rates.

\subsection{Quantum kinetic equations: summary}

Having confirmed the validity of the kinetic equations via two
independent but complementary methods, we now summarize the quantum
kinetic equations in  a form amenable to numerical study. For this
purpose it is convenient to define the hermitian combinations \be
n_R(t) = n_{12}(t)+n_{21}(t) ~;~ n_I(t) = i(n_{12}(t)-n_{21}(t))
\label{combos}\ee in terms of which the quantum kinetic equations
for the distribution functions and coherences become (suppressing
the momentum label)

\bea \dot{n}_{11}  & = &
-\Gamma_{aa}\cos^2\tm\Big[n_{11}-n_{eq,1}\Big] - \frac{
\Gamma_{aa}}{4}\sin 2\tm~ n_R \label{dotn11fin}
\\ \dot{n}_{22}  & = &  -\Gamma_{aa}\sin^2\tm \Big[n_{22}-n_{eq,2}\Big] - \frac{
\Gamma_{aa}}{4}\sin 2\tm ~ n_R \label{dotn22fin}\\\dot{n}_{R} & = &
-(\omega_2-\omega_1) ~ n_I-\frac{\Gamma_{aa}}{2} n_{R} - \frac{
\Gamma_{aa}}{2}\sin 2\tm \Big[(n_{11}-n_{eq,1}) + (n_{22}-n_{eq,2})
\Big] \label{dotn12fin}\\ \dot{n}_{I} & = &   (\omega_2-\omega_1) ~
n_R -\frac{\Gamma_{aa}}{2} n_{I}   \,,\label{dotn21fin}\eea with the
active and sterile distribution functions related to the quantities
above as follows \bea N_a(k;t) & = & \cos^2\tm n_{11}(k;t)+\sin^2\tm
n_{22}(k;t)+ \frac{1}{2}\sin 2\tm n_{R}(k;t) \label{Nat} \\N_s(k;t)
& = & \sin^2\tm n_{11}(k;t)+\cos^2\tm n_{22}(k;t)- \frac{1}{2}\sin
2\tm n_{R}(k;t)\label{Nsat} \,.   \eea

In the perturbative limit when $\Gamma_{aa}\sin2\tm/\Delta \omega
\ll 1$ which as argued above is the correct limit in all but for a
possible small region near an MSW resonance\cite{hozeno},   the set
of kinetic equations simplify to

\bea \dot{n}_{11}  & = &
-\Gamma_{aa}\cos^2\tm\Big[n_{11}-n_{eq,1}\Big]   \label{dotn11fin2}
\\ \dot{n}_{22}  & = &  -\Gamma_{aa}\sin^2\tm \Big[n_{22}-n_{eq,2}\Big]  \label{dotn22fin2}\\\dot{n}_{R} & = &
-\Big(\omega_2-\omega_1\Big) ~ n_I-\frac{\Gamma_{aa}}{2} n_{R}   \label{dotn12fin2}\\
\dot{n}_{I} & = &   \Big(\omega_2-\omega_1\Big) ~ n_R
-\frac{\Gamma_{aa}}{2} n_{I} \,.\label{dotn21fin2}\eea In this case
the active and sterile populations are given by (suppressing the
momentum variable)

\bea N_a(t) & = &
\cos^2(\tm)\Bigg[n_{eq}(\omega_1)+\Big(n_{11}(0)-n_{eq}(\omega_1)\Big)\,e^{-\Gamma_1
t}\Bigg] \nonumber \\ & + &
\sin^2(\tm)\Bigg[n_{eq}(\omega_2)+\Big(n_{22}(0)-n_{eq}(\omega_2)\Big)\,e^{-\Gamma_2
t}\Bigg] \nonumber \\ &+&  ~\sin(2\tm) \,e^{-\frac{\Gamma_{aa}}{2}
t}\,\cos\left[(\omega_1-\omega_2)t\right]~
 n_{12}(0) \,.\label{numberactfin}\eea

\bea N_s(t) & = &
\sin^2(\tm)\Bigg[n_{eq}(\omega_1)+\Big(n_{11}(0)-n_{eq}(\omega_1)\Big)\,e^{-\Gamma_1
t}\Bigg] \nonumber \\ & + &
\cos^2(\tm)\Bigg[n_{eq}(\omega_2)+\Big(n_{22}(0)-n_{eq}(\omega_2)\Big)\,e^{-\Gamma_2
t}\Bigg] \nonumber \\ &-&  ~\sin(2\tm) \,e^{-\frac{\Gamma_{aa}}{2}
t}\,\cos\left[(\omega_1-\omega_2)t\right]~
 n_{12}(0) \,,\label{numberstefin}\eea where \be \Gamma_1(k) =
 \Gamma_{aa}(k) \cos^2\tm ~~;~~\Gamma_2(k) = \Gamma_{aa}(k)
 \sin^2\tm \ee and assumed that $n_{12}(0)$ is real as is the case
  when  the initial density matrix is  diagonal both in the flavor
  or $1,2$ basis, \be n_{12}(0) = \Bigg\{\begin{array}{ll}
                                     \frac{1}{2} \sin 2\tm ~
(N_a(0)-N_s(0)) ~~\mathrm{diagonal~in ~flavor~basis}  \\
                                    0 ~~~~~~~~~~~~~~~~~~~~~~~~~~~~~~~~~~~\mathrm{diagonal~ in
                                    ~1,2~basis}
                                  \end{array} \label{inival12} \ee

It is clear that the evolution of the active and sterile
distribution functions \emph{cannot}, in general,  be written as
simple rate equations.

From the expressions given above for the quantum kinetic equations
it is straightforward to generalize to account for the fermionic
nature of neutrinos: the equilibrium distribution functions are
replaced by the Fermi-Dirac distributions, and Pauli blocking
effects enter in the explicit calculation of the damping rates.

\section{Transition probabilities and coherences}\label{sec:transprob}

\subsection{ A ``transition probability'' in a
medium}\label{subsec:prob}

 The
concept of a transition probability as typically used in neutrino
oscillations is not suitable   in a medium when the description is
not in terms of wave functions but density matrices. However, an
equivalent concept can be provided as follows. Consider expanding
the active and sterile fields in terms of creation and annihilation
operators. In the ultrarelativistic limit the positive frequency
components are obtained from the relation (\ref{as})and their
ensemble averages in the reduced density matrix are given by \be
\varphi_{a,s}(\vk,t) \equiv \langle a_{a,s}(\vk,t)\rangle \,.\ee The
kinetic equations for $\langle a_{1,2} (\vk ) \rangle (t)$ are found
to be \bea \frac{d}{dt} \langle a_{1} (\vk ) \rangle(t) =
\Big(-i\omega_1(k) - \frac{\Gamma_1(k)}{2}\Big) \langle a_{1} (\vk )
\rangle (t) - \frac{\widetilde{\Gamma}}{2}~ \langle a_{2} (\vk )
\rangle(t) \label{a1dot}\\\frac{d}{dt} \langle a_{2} (\vk )
\rangle(t) = \Big(-i\omega_2(k) - \frac{\Gamma_2(k)}{2}\Big) \langle
a_{2} (\vk,t) \rangle - \frac{\widetilde{\Gamma}}{2} ~ \langle a_{1}
(\vk ) \rangle(t)\, , \label{a2dot}\eea  where $\widetilde{\Gamma}$
has been defined in eqn. (\ref{gammas}). To leading order in
$\widetilde{\Gamma}/\Delta \omega$ the solutions of these kinetic
equations are \bea \langle a_{1} (\vk ) \rangle(t) & = & \langle
a_{1} (\vk ) \rangle(0) ~e^{-i\omega_1t} e^{-\frac{\Gamma_1}{2}t}-
\frac{i\widetilde{\Gamma}}{2\Delta\omega} \langle a_{2} (\vk )
\rangle(0) \Big[e^{-i\omega_2t} e^{-\frac{\Gamma_2}{2}t}  -
e^{-i\omega_1t} e^{-\frac{\Gamma_1}{2}t}\Big] \label{a1oft}\\\langle
a_{2} (\vk ) \rangle(t) & = &   \langle a_{2} (\vk )
\rangle(0)~e^{-i\omega_2t}
e^{-\frac{\Gamma_2}{2}t}+\frac{i\widetilde{\Gamma}}{2\Delta\omega}
\langle a_{1} (\vk ) \rangle(0) \Big[e^{-i\omega_1t}
e^{-\frac{\Gamma_1}{2}t} - e^{-i\omega_2t}
e^{-\frac{\Gamma_2}{2}t}\Big] \label{a2oft} \,.\eea The initial
values $\langle a_{1,2} (\vk ) \rangle(0)$ determine the initial
values   $\varphi_{a,s}(\vk;0)$, or alternatively, giving the
initial values $\varphi_{a,s}(\vk;0)$ determines $\langle a_{1,2}
(\vk ) \rangle(0)$. Consider the case in which the initial density
matrix is such that \be    \langle a_{a}(\vk,0) \rangle \equiv
\varphi_a(\vk)\neq 0~~;~~ \langle a_{s}(\vk,0)\rangle \equiv
\varphi_s(\vk,0) =0 \label{inifi}\ee   the initial values of
$\langle a_{1,2} (\vk ) \rangle(0)$ are obtained   by inverting the
relation (\ref{fia}) from which we find \be \varphi_s(\vk,t) =
-\frac{1}{2}\sin2\tm (1-i {\widetilde{\gamma}})\Big[e^{-i\omega_1t}
e^{-\frac{\Gamma_1}{2}t}-e^{-i\omega_2t} e^{-\frac{\Gamma_2}{2}t}
\Big]\varphi_a(\vk,0)\,, \label{fisteoft}\ee this result coincides
with that found in ref.\cite{hobos}. We can interpret the
``transition probability'' as \be P_{a\rightarrow s}(\vk,t)= \Bigg|
\frac{\varphi_s(\vk,t)}{\varphi_a(\vk,0)}\Bigg|^2 =
\frac{1}{4}\sin^2 2\tm ~
\Bigg[e^{-\Gamma_1t}+e^{-\Gamma_2t}-2\cos\Big(\big( \omega_2
-\omega_1\big) t\Big)~e^{-\frac{1}{2}(\Gamma_1+\Gamma_2)t}\Bigg]
\label{Pas}\ee where we have neglected perturbative corrections of
$\mathcal{O}(\widetilde{\gamma})$. This result  coincides with that
obtain in ref.\cite{hobos} from the effective action, and confirms a
similar result for neutrinos with standard model
interactions\cite{hozeno}. We emphasize that this ``transition
probability'' is \emph{not} obtained from the time evolution of
single particle wave functions, but from \emph{ensemble averages in
the reduced density matrix}: the initial density matrix features a
non-vanishing expectation value of the active field but a vanishing
expectation value of the sterile field, however, upon time evolution
the density matrix develops an expectation value of the sterile
field. The relation between the transition probability (\ref{Pas})
and the time evolution of the distribution functions and coherences
is now explicit, the first two terms in (\ref{Pas}) precisely
reflect the time evolution of the distribution functions
$n_{11},n_{22}$ with time scales $1/\Gamma_{1,2}$ respectively,
while the last, oscillatory term is the interference between the
active and sterile components and is damped out on the decoherence
time scale $\tau_{dec}$. This analysis thus confirms the results in
ref.\cite{hozeno}.

\subsection{Coherences}\label{subsec:coherence}

The time evolution of the off-diagonal coherence $\langle
a^\dagger_1(\vk) a_2(\vk) \rangle (t) $ is determined by the kinetic
equation (\ref{dotn12}), neglecting perturbatively small corrections
of $\mathcal{O}(\widetilde{\gamma})$ \be \langle a^\dagger_1(\vk)
a_2(\vk) \rangle (t) = \langle a^\dagger_1(\vk) a_2(\vk) \rangle
(0)~e^{i\Delta \omega t } e^{-\frac{\Gamma_{aa}}{2}t}
\label{ofdia}\ee where we have used the relations (\ref{gammas} ) in
the ultrarelativistic limit. Therefore, in perturbation theory, if
the initial density matrix is off-diagonal in the $1-2$ basis
(propagating modes in the medium) the off-diagonal correlations are
exponentially damped out on the coherence time scale $\tau_{dec} =
2/\Gamma_{aa}(k)$. This coherence term  and its hermitian conjugate
are precisely the ones responsible for the oscillatory term in the
transition probability (\ref{Pas}).  An important consequence of the
damping of the off-diagonal coherences is that  in perturbation
theory the equilibrium density matrix is \emph{diagonal  in the
basis of the propagating modes in the medium}. This result confirms
the arguments in ref.\cite{hochar}. As can be seen from the
expression of the transition probability (\ref{Pas}) this is
precisely the time scale for suppression of the \emph{oscillatory}
interference term. However, the transition probability is \emph{not}
suppressed on this coherence time scale, the first two terms in
(\ref{Pas}) reflect the fact that the \emph{occupation} numbers
build up  on     time scales $1/\Gamma_1;1/\Gamma_2$ respectively
and the interference term is  exponentially suppressed on the
decoherence time scale $\tau_{dec}=2/(\Gamma_1+\Gamma_2)$. For small
mixing angle in the medium $\tm$ all of these time scales can be
widely different.

It is noteworthy to compare the transition probability
(\ref{fisteoft}) with the distribution functions
(\ref{numberactfin},\ref{numberstefin}). The first two,
non-oscillatory terms in (\ref{fisteoft}) describe the same time
evolution as the distribution functions $n_{11},n_{22}$ of the
propagating modes in the medium, while the last, oscillatory term
describes the interference between these. This confirms the results
and arguments provided in ref.\cite{hozeno}.

\section{From the quantum Master equation to the QKE for the ``polarization''
vector}\label{sec:polar}

The results of the previous section allows us to establish a
correspondence between the quantum master equation (\ref{qmefin})
the quantum kinetic equations (\ref{dotn11fin}-\ref{dotn21fin}) and
the quantum kinetic equation for a polarization vector often used in
the literature\cite{mckellar,wong}.  Following ref.\cite{bohot}, let
us define the ``polarization vector'' with the following components,

\bea P_0(\vk,t) & = & \langle a^\dagger_a(\vk,t)a_a(\vk,t)+
a^\dagger_s(\vk,t)a_s(\vk,t) \rangle =
N_a(k,t)+N_s(k,t)\label{P0}\\P_x(\vk,t) & = & \langle
a^\dagger_a(\vk,t)a_s(\vk,t)+ a^\dagger_s(\vk,t)a_a(\vk,t) \rangle
\label{Px}\\P_y(\vk,t) & = & -i\langle
a^\dagger_a(\vk,t)a_s(\vk,t)-a^\dagger_s(\vk,t)a_a(\vk,t) \rangle
\label{Py}\\P_z(\vk,t) & = & \langle a^\dagger_a(\vk,t)a_a(\vk,t)-
a^\dagger_s(\vk,t)a_s(\vk,t) \rangle = N_a(k,t)-N_s(k,t)
\label{Pz}\eea where the creation and annihilation operators for the
active and sterile fields are related to those that create and
annihilate the propagating modes in the medium $1,2$ by eqn.
(\ref{as}), and the angular brackets denote expectation values in
the reduced density matrix $\rho_{S,i}$ which obeys the quantum
master equation (\ref{qmefin}). In terms of the population and
coherences $n_{ij}$ the elements of the polarization vector are
given by

\bea P_0 & = & n_{11}+n_{22} \label{P01} \\
P_x & = & -\sin2\tm \Big(n_{11}-n_{22}\Big)+\cos2\tm n_R
\label{Px1}\\ P_y & = & - n_I \label{Py1}\\P_z & = & \cos2\tm
\Big(n_{11}-n_{22}\Big)+ \sin2\tm n_R \label{Pz1}\eea where
$n_{R,I}$ are defined by equation (\ref{combos}). Using the quantum
kinetic equations (\ref{dotn11fin}-\ref{dotn21fin}) we find \be
\frac{dP_0}{dt} = -\frac{\Gamma_{aa}}{2}P_z -
\frac{\Gamma_{aa}}{2}\Bigg[\Big(n_{11}-n_{eq,1}\Big)+
\Big(n_{22}-n_{eq,2}\Big) \Bigg]+\frac{\Gamma_{aa}}{2}\cos2\tm
\Big(n_{eq,1}-n_{eq,2}\Big)\label{dotP0}\ee \be \frac{dP_x}{dt} =
-(\omega_2-\omega_1) \cos2\tm n_I - \frac{\Gamma_{aa}}{2}P_x -
\frac{\Gamma_{aa}}{2}\sin2\tm \Big(n_{eq,1}-n_{eq,2}\Big)
\label{dotPx}\ee \be \frac{dP_y}{dt} = -(\omega_2-\omega_1)n_R
-\frac{\Gamma_{aa}}{2}P_y \label{dotPy}\ee \be \frac{dP_z}{dt} =
-(\omega_2-\omega_1) \sin2\tm n_I -\frac{\Gamma_{aa}}{2}P_z -
\frac{\Gamma_{aa}}{2}\Bigg[\Big(n_{11}-n_{eq,1}\Big)+
\Big(n_{22}-n_{eq,2}\Big) \Bigg] \label{dotPz}\ee

We now \emph{approximate}\be \Big(n_{eq,1}-n_{eq,2}\Big)  \sim
\frac{(\omega_2-\omega_1)}{T} ~n'_{eq}(x) \sim 0 \,,
\label{eqmass}\ee thus neglecting the last terms in eqns.
(\ref{dotP0},\ref{dotPx}), introducing the vector $\vec{V}$ with
components \be \vec{V} = (\omega_2-\omega_1)~\Big(\sin 2\tm, 0 ,
-\cos2\tm\Big) \label{VecV}\ee   we find the following equations of
motion for the polarization vector \be \frac{d\vec{P}}{dt} =
\vec{V}\times\vec{P} - \frac{\Gamma_{aa}}{2}\Big(P_x \hat{x}+P_y
\hat{y}\Big)+ \frac{dP_0}{dt}\hat{z} \label{QKEpol}\ee This equation
is exactly of the form \be \frac{d\vec{P}}{dt} =
\vec{V}\times\vec{P} - D \vec{P}_T + \frac{dP_0}{dt}\hat{z}
\label{QKEpol2}\ee used in the
literature\cite{stodolsky,mckellar,wong,foot,dibari}, where $D$ and
$\vec{P}_T$ can be  identified from eqn. (\ref{QKEpol}).

Therefore the quantum kinetic equation for the polarization vector
(\ref{QKEpol}) is \emph{equivalent} to the full set of quantum
kinetic equations (\ref{dotn11fin}-\ref{dotn21fin}) or equivalently
to equations (\ref{dotn11}-\ref{dotn21}) under the approximation
(\ref{eqmass}). Furthermore since the quantum kinetic equations
(\ref{dotn11fin}-\ref{dotn21fin}) have been proven to be equivalent
to the time evolution obtained from the effective action, we
conclude that the kinetic equation for the polarization vector
(\ref{QKEpol2}) is completely equivalent to the effective action and
the quantum master equation under the approximations discussed
above. This equivalence between the effective action, the kinetic
equations obtained from quantum Master equation and the kinetic
equations for the polarization vector makes explicit  that the
\emph{fundamental} scales for decoherence and damping are determined
by  $\Gamma_{1,2}$, which are twice the damping rates of the
quasiparticle modes. These are completely determined by the complex
poles of the propagator in the medium. Furthermore the formulation
in terms of the effective action, or equivalently the quantum master
equation (\ref{qmefin}) provides more information: for example from
both we can extract the transition probability $P_{a\rightarrow s} $
in the medium from expectation values of the field operators (or
creation/annihilation operators) in the reduced density matrix,
leading unequivocally to the expression (\ref{Pas}) which indeed
features the \emph{two} relevant time scales. Furthermore  it
directly yields information on the off-diagonal coherences
(\ref{ofdia}) which fall off on the decoherence time scale
$\tau_{dec}=2/\Gamma_{aa}$, thus elucidating that the reduced
density matrix in equilibrium (the asymptotic long time limit) is
\emph{diagonal in the 1-2 basis}. While this information could be
extracted from linear combinations of $P_x,P_y$ it is hidden in the
solution of the kinetic equation for the polarization, whereas it is
exhibited clearly in the quantum kinetic equations
(\ref{dotn11}-\ref{dotn21}) in the regime in which perturbation
theory is applicable $|\Gamma_{aa}\sin2\tm/(\omega_2-\omega_1)|\ll1$
. In this regime, which as argued above is the most relevant, the
  set of quantum kinetic equations (\ref{dotn11fin2}-\ref{dotn21fin2}) combined with the relations
  (\ref{nacti}-\ref{stenu})  yield  a
  much simpler and  numerically amenable description of the time evolution of
  the populations and coherences: the active and sterile
  distribution functions are given by equations
  (\ref{numberactfin},\ref{numberstefin}) and the off-diagonal coherence by eqn.
  (\ref{ofdia}).
Therefore, while the kinetic equation for the polarization and the
quantum kinetic equations (\ref{dotn11fin2}-\ref{dotn21fin2}) are
equivalent and both are fundamentally consequences of the effective
action or equivalently the quantum master equation, the study of
sterile neutrino production in the early Universe does \emph{not}
implement any of these equivalent quantum kinetic formulations but
instead assume a phenomenological approximate description in terms
of a simple \emph{rate equation}\cite{foot,kev1,dibari}, which
implies only one damping scale. Such a simple  rate equation
\emph{cannot} describe accurately the time   evolution of
distribution functions and coherences which involve  two different
time scales (away from MSW resonances). In our view, part of the
problem in this formulation is the time averaged transition
probability introduced in ref.\cite{foot} which inputs the usual
quantum mechanical vacuum transition probability but \emph{damped
by a simple exponential on the decoherence time scale}, clearly in
contradiction with the result (\ref{Pas}) obtained from the
\emph{reduced quantum density matrix}. Within the kinetic
formulation for the time evolution of the polarization vector
$P_0,\vec{P}$, eqn. (\ref{QKEpol}) it is not possible to extract the
notion of a transition probability because the components of
polarization vector are expectation values of \emph{bilinear}
operators in the reduced density matrix. Instead, the concept of
active-sterile transition probability \emph{can} be established in a
medium via expectation values of the field operators (or their
creation/annihilation operators) in the reduced density matrix es
discussed in section (\ref{subsec:prob}).

\section{Conclusions}\label{sec:conclu}

Our goal is to study the non-equilibrium quantum kinetics of
production of active and sterile neutrinos in a medium. We make
progress towards that goal by studying a model of an active and a
sterile \emph{mesons} coupled to a bath in thermal equilibrium via
couplings that model charged and current interactions of neutrinos.
The dynamical aspects of mixing, oscillations, decoherence and
damping are fairly robust and the results of the study can be simply
modified to account for Pauli blocking effects of fermions and the
detailed form of the  matter potential. As already discussed in
ref\cite{hobos} with simple modifications, such as the detailed form
of $V_{aa}$ including the CP-odd and even terms\cite{notzold}, and
the Fermi-Dirac distributions for the equilibrium ones, this model
provides a remarkably faithful description of the non-equilibrium
dynamics of neutrinos.

 We obtained the quantum kinetic equations for the active and
 sterile  species via two independent but complementary methods. The
 first method obtains the non-equilibrium effective action for the
 active and sterile species after integrating out the bath degrees
 of freedom. This description provides a non-perturbative Dyson-like
 resummation of the self-energy radiative corrections, and the dynamics of the
 distribution functions is completely determined by the solutions of
 a Langevin equation with a noise term that obeys a generalized
 fluctuation-dissipation relation. The important ingredient in this
 description is the \emph{full propagator}. The poles of the
 propagator correspond to two quasiparticle modes whose frequencies obey
 the usual dispersion relations of neutrinos in a medium with the
 corrections from the index of refraction (forward scattering), with
  damping rates (widths)  \be \Gamma_1
 =\Gamma_{aa}\cos^2\tm~;~\Gamma_2 = \Gamma_{aa}\sin^2\tm \,.\ee where
 $\Gamma_{aa}$ is the interaction rate of the active species
 \emph{in absence of mixing} (in the ultrarelativistic limit) and
 $\tm$ is the mixing angle \emph{in the medium}. These \emph{two
 damping} scales, along with the quasiparticle frequencies completely
  determine the evolution of the  distribution functions. This is
  one of the important aspects of the kinetic description in terms
  of the non-equilibrium effective action:\emph{ the dispersion relations
  and damping rates of the quasiparticle modes corresponding to the
  poles of the full propagator completely determine the
  non-equilibrium evolution of the distribution functions and
  coherences.}

  We also obtained the quantum master equation for the
  \emph{reduced} density matrix for the ``neutrino degrees of
  freedom'' by integrating (tracing) over the bath degrees of
  freedom taken to be in thermal equilibrium. An important aspect of
  the derivation consists in including the matter potential, or
  index of refraction from forward scattering to lowest order in the
  interactions in the unperturbed Hamiltonian. This method provides
  a re-arrangement of the perturbative expansion that includes
  self-consistently the index of refraction corrections and builds in  the
  correct propagation frequencies in the medium. In this manner the
   the reduced density matrix (in the interaction
  picture ) evolves in time \emph{only through second order
  processes}. From the reduced density matrix we obtain the quantum
  kinetic equations for the distribution functions and coherences.
  These are \emph{exactly the same as those obtained from the
  non-equilibrium effective action}. We also obtain the kinetic
  equation for coherences and introduce a generalization of the
  active-sterile transition probability by obtaining the time
  evolution of expectation values of the active and sterile
  \emph{fields} in the reduced quantum density matrix. Within the
  realm of validity of the perturbative expansion the set of kinetic
  equations for the distribution functions and coherences are given
  by

\bea \dot{n}_{11}   & = & -\Gamma_{aa}\cos^2\tm
\Big[n_{11}-n_{eq,1}\Big] ~~;~~
  \dot{n}_{22}    =    -\Gamma_{aa}\sin^2\tm\Big[n_{22}-n_{eq,2}\Big]   \nonumber \\
  \dot{n}_{12} &  =  &  \Big[-i\Big(\omega_2(k)-\omega_1(k)\Big)
-\frac{\Gamma_{aa}}{2}\Big]n_{12} ~~;~~   {n}_{21}   = n^*_{12} \,,
\label{quankine} \eea where   $ n_{eq,j} = n(\omega_j(k))$ are the
equilibrium distribution functions for the corresponding propagating
modes, $ \omega_{1,2}(k)$ are the dispersion relations in the medium
including the index of refraction,  and the active and sterile
distribution functions are given by

 \bea   N_a(k;t)  & =  &
\cos^2\tm n_{11}(k;t)+\sin^2\tm n_{22}(k;t)+ \frac{1}{2}\sin 2\tm
\Big(n_{12}(k;t)+n_{21}(k;t)\Big) \label{relas1} \\N_s(k;t)  & =  &
\sin^2\tm n_{11}(k;t)+\cos^2\tm n_{22}(k;t)- \frac{1}{2}\sin 2\tm
\Big(n_{12}(k;t)+n_{21}(k;t)\Big)\label{relas2} \,.  \eea

The set of equations (\ref{quankine})  provide a simple system of
uncoupled rate equations amenable to numerical study, whose solution
yields the active and sterile distribution functions via the
relations (\ref{relas1},\ref{relas2}), with straightforward
modifications for fermions.

From the kinetic equations above, it is found that the  coherences
  \be n_{12} = \langle a^\dagger_1 a_2 \rangle \ee which are off-diagonal (in
  the $1-2$ basis of propagating modes in the medium) expectation values in the reduced quantum
  density matrix     are
  exponentially suppressed on a \emph{decoherence} time scale
  $\tau_{dec} = 2/\Gamma_{aa}$ indicating that the equilibrium reduced density
  matrix is diagonal in the $1-2$ basis, confirming the arguments in
  ref.\cite{hochar}.

 The
generalization of the active-sterile transition probability in the
medium via the \emph{expectation value of the active and sterile
fields in the reduced quantum density matrix} yields

\be P_{a\rightarrow s} = \frac{1}{4}\sin^2 2\tm ~
\Bigg[e^{-\Gamma_1t}+e^{-\Gamma_2t}-2\cos\Big(\big( \omega_2
-\omega_1\big) t\Big)~e^{-\frac{\Gamma_{aa}}{2}t}\Bigg]\ee  this
result shows that the active-sterile transition probability depends
on the \emph{two damping time scales of the quasiparticle modes in
the medium} which are also the time scales of kinetic evolution of
the distribution functions, and confirms the results of
refs.\cite{hozeno}.

Finally, from the full set of quantum kinetic equations
(\ref{dotn11fin2}-\ref{dotn21fin2})  and the approximation
(\ref{eqmass}) we have obtained the set of quantum kinetic equations
for   the polarization vector, most often used in the literature,
\be\label{poleq} \frac{d\vec{P}}{dt} = \vec{V}\times\vec{P} -
\frac{\Gamma_{aa}}{2}\Big(P_x \hat{x}+P_y \hat{y}\Big)+
\frac{dP_0}{dt}\hat{z}  \ee where the relation between the
components of the polarization vector $P_0,\vec{P}$ and the
distribution functions and coherences is explicitly given by eqns.
(\ref{P0}-\ref{Pz}) (or equivalently (\ref{P01}-\ref{Pz1})), and
$\vec{V}$ is given by eqn. (\ref{VecV}). Thus we have unambiguously
established the direct relations between the effective action,
quantum master equation, the full set of kinetic equations for
population and coherences and the quantum kinetic equations in terms
of the ``polarization vector'' most often used in the literature.
These are all equivalent, but the effective action approach
distinctly shows that the \emph{two} independent fundamental damping
scales are those associated with $\Gamma_{1,2}$, namely the damping
rates of the quasiparticles in the medium, which are determined by
the complex poles of the propagator. Furthermore in the regime of
validity of perturbation theory, the set of kinetic equations
(\ref{quankine}) obtained from the quantum master equation yield  a
simple and clear understanding of the different time scales for the
active and sterile distribution functions and a remarkably concise
description of active and sterile production when combined with the
relations (\ref{relas1},\ref{relas2}). These simpler set of rate
equations are hidden in the kinetic equation (\ref{poleq}).

We have also argued that the simple phenomenological rate equation
used in numerical studies of sterile neutrino production in the
early Universe is \emph{not} an accurate description of the
non-equilibrium evolution, and trace its shortcomings to the time
integral of an overly simplified  description of the transition
probability in the medium.

Our study focused on a \emph{scalar model} that features many
similarities to but also distinct differences with the theory of
mixed neutrinos. The dispersion relations, medium dependence of the
mixing angles, transition probabilities for ensemble averages, and
dependence of the damping rates of the propagating modes on the
active collision rate are robust features in common with the case of
neutrinos. These similarities are strengthened by the fact that the
kinetic equations obtained in this article are \emph{identical} to
those available in the literature in terms of the polarization
vector, with the bonus that we provide a different interpretation
that highlights the role of the non-equilibrium evolution in terms
of the physical propagating modes. All of these similarities and the
combination of results obtained in this study and those reported
in\cite{hobos,hozeno} lend   support to the expectation that the
results obtained in this study are relevant for the description of
the kinetics of neutrinos.

There are, however, differences with the neutrino case that must
eventually be addressed for a more complete treatment and
understanding: spinorial and chiral structures, although these are
not directly accounted for either in the quantum mechanical
description of neutrino oscillations nor in the phenomenological
description of the kinetics, Fermionic nature of the neutrino field,
which enters in the distribution function, however, the simplicity
of the kinetic equations found in this article allow a simple
replacement of the distribution functions by the Fermi-Dirac one,
automatically including Pauli blocking, furthermore, the matter
potential in the case of neutrinos features both a CP-odd term
arising from the lepton and baryon asymmetry, and a CP-even term
that depends solely on temperature, the overall sign of the matter
potential is determined by these two contributions. For the case of
sterile neutrinos with $keV$ masses, an MSW resonance is only
available when the CP-odd term dominates. Our study in this article
is general, without specifying a particular form of the matter
potential and  addressed all possibilities with or without MSW
resonances. The only specific aspect is that the matter potential is
flavor diagonal and only features an entry in the active-active
matrix element.

While the model studied here is clearly a simplification of the case
of neutrinos, the body of results and similarities established with
the neutrino case \emph{suggest} a reliable description of the
quantum kinetics.  A more detailed study of the impact of the
differences on the non-equilibrium dynamics will be the subject of
forthcoming work.

\begin{acknowledgments}   The authors  acknowledge support from the US NSF
under grants PHY-0242134, 0553418. C.M.Ho acknowledges partial
support through the Andrew Mellon Foundation and the Zaccheus Daniel
Fellowship.
\end{acknowledgments}

\appendix

\section{A simpler case}\label{app:num}

Consider for simplicity the case of one scalar field. The solution
of the Langevin equation is given by \begin{equation} \Psi_ {\vk}(t
) = \dot{g}(k;t)\,\Psi^0_{\vec k} + {g} (k;t)\,
 \Pi^0_{ \vec k} + \int^t_0
g (k;t')~\xi_{ \vec k}(t-t') dt' \,, \label{inhosolutionap}
\end{equation} where the dot stands for derivative with respect to
time. In the Breit-Wigner approximation and setting $Z_k=1$  \be
g(k;t) =
\frac{\sin[\Omega(k)\,t]}{\Omega(k)}\,e^{-\frac{\Gamma(k)}{2} t}\,.
\label{gs}\ee where $\Omega(k)$ is the position of the quasiparticle
pole (dispersion relation)   and its width is given by  \be
\Gamma(k) = \frac{\Sigma_I(\Omega(k))}{\Omega(k)}\label{widths}\ee
The particle number is given by \be N(k,t) = \frac{1}{2W(k)}
\Bigg[\langle \dot{\Psi}(\vk,t)\dot{\Psi}(-\vk,t) \rangle + W^2(k)
\langle {\Psi}(\vk,t) {\Psi}(-\vk,t) \rangle \Bigg]- \frac{1}{2}
\label{parnum}\ee where $W(k)$ is the bare frequency. Taking the
initial density matrix of the field $\Psi$ to be that corresponding
to a free-field with arbitrary non-equilibrium initial distribution
function $N(k;0)$ and carrying out both averages, over the initial
density matrix for the field and of the quantum noise and using that
the average of the latter vanishes, we find \be
N(k;t)=N_1(k;t)+N_2(k;t)-\frac{1}{2}\label{nofkt}\ee with \bea
N_1(k;t)  & = &
\frac{1+2N(k;0)}{4\,W^2(k)}\Bigg[(\ddot{g}(k;t))^2+2W^2(k)(\dot{g}(k;t))^2
+W^4(k)g^2(k;t)\Bigg]\label{N1}  \\ N_2(k;t) & = &\frac{1}{2W^2(k)}
\int
\frac{d\omega}{2\pi}\,\Sigma_I(k;\omega)\left[1+2n(\omega)\right]\,
\Bigg[W^2(k)\big|h(\omega,t)\big|^2+\big|f(\omega,t)\big|^2 \Bigg]
 \label{N2} \eea where \bea
\\ h(\omega,t) & = & \int_0^t e^{-i\omega t'} g(k;t')dt' \label{h} \\
f(\omega,t) & = & \int_0^t e^{-i\omega t'} \dot{g}(k;t')dt'
\label{f}\eea The terms $N_1(k;t);N_2(k;t)$ have very different
origins: the term $N_1(k;t)$ depends on the initial condition and
originates in the first two terms in (\ref{inhosolution}) namely
those \emph{independent of the noise}, which survive upon taking the
average over the noise. The term $N_2(k;t)$ is independent of the
initial conditions and is solely determined by the correlation
function of the noise term and is a consequence of the fluctuation
dissipation relation. Using the expression (\ref{gs}) we find \be
N_1(k;t) =   \left[\frac{1}{2}+ N(k;0)\right]\,e^{-\Gamma(k)
t}\left[1+\sin^2(\Omega(k) t) \left(\frac{\Omega^2(k)-W^2(k)}{2W
(k)\Omega (k)}\right)^2
+\mathcal{O}\left(\frac{\Gamma^2(k)}{\Omega^2(k)}\right)\right]\label{N1a}\ee
where the neglected terms of order $\Gamma^2(k)/\Omega^2(k) \ll 1$
are perturbatively small. The oscillatory term in (\ref{N1a})
averages out on a short time scale $1/\Omega(k) \ll 1/\Gamma(k)$ and
we can replace (\ref{N1a}) by its   average over this short time
scale yielding \be N_1(k;t) \approx  \left[\frac{1}{2}+
N(k;0)\right]\,e^{-\Gamma(k) t}\left[1+\frac{1}{2}
\left(\frac{\Omega^2(k)-W^2(k)}{2W (k)\Omega (k)}\right)^2
+\mathcal{O}\left(\frac{\Gamma^2(k)}{\Omega^2(k)}\right)\right]\label{N1b}\ee

In perturbation theory   $\Omega^2(k)-W^2(k)/2W (k)\Omega (k) \ll
1$, can   be neglected to leading order in perturbative quantities,
thus we obtain

\be N_1(k;t) \approx \left[\frac{1}{2}+ N(k;0)\right]\,e^{-\Gamma(k)
t}\,. \label{N1fin}\ee Using the fact that $\Sigma_I(\omega) =
-\Sigma_I(-\omega)$ we can perform the integrals in $N_2(k;t)$ in
the narrow width (Breit-Wigner) approximation by using eqn.
(\ref{widths}), with the result

\be  N_2(k;t)   \simeq Z^{\,2}_k
\left[\frac{W^2(k)+\Omega^2(k)}{2W(k)\Omega(k)}\right]\,\left[\frac{1}{2}+n(\Omega(k))\right]\,
\left(1-e^{-\Gamma(k)t}\right)+\mathcal{O}\left(\frac{\Gamma^2(k)}{\Omega^2(k)}\right)\,
\ee Replacing in perturbation theory \be Z_k \approx 1 ~~;~~
\frac{W^2(k)+\Omega^2(k)}{2W(k)\Omega(k)} \approx 1 \ee we find \be
N(k;t)= n(\Omega(k))+
\left(N(k;0)-n(\Omega(k))\right)\,e^{-\Gamma(k)t} \label{finnum}\ee
which is the solution of the usual kinetic equation \be \frac{d
N(k;t)}{dt} = -\Gamma(k) \left(N(k;t)-N_{eq}(k)\right)
\label{kinesim}\ee where \be N_{eq}(k) = n(\Omega(k))\ee

It is important to highlight the series of approximations that led
to this result: i) the narrow width (Breit-Wigner) approximation,
ii) $Z_k\sim 1$, iii) $\Omega^2(k) \sim W^2(k)$, iv)
$\Gamma(k)/\Omega(k) \ll1$, these approximations are all warranted
in perturbation theory. Clearly including perturbative corrections
lead to perturbative departures of the usual kinetic equation and of
the equilibrium distribution function.

\section{Quantum Master equation }\label{app:qme}

Taking the trace over the bath variables with the  factorized
density matrix (\ref{fact}), the double commutator in equation
(\ref{2ndordred}) becomes \bea -\sum_{\vk} \int^t_0 dt' && \Big\{
\phi_a(t)\phi_a(t')\rho_{S,i}(t)~\mathrm{Tr}_B\rho_B(0)
\mathcal{O}(t)\mathcal{O}(t') \nonumber \\ && +
\rho_{S,i}(t)\phi_a(t')\phi_a(t) ~\mathrm{Tr}_B\rho_B(0)
\mathcal{O}(t')\mathcal{O}(t) \nonumber \\&& -
\phi_a(t)\rho_{S,i}(t)\phi_a(t') ~\mathrm{Tr}_B\rho_B(0)
\mathcal{O}(t')\mathcal{O}(t) \nonumber \\&& -
\phi_a(t')\rho_{S,i}(t)\phi_a(t) ~\mathrm{Tr}_B\rho_B(0)
\mathcal{O}(t)\mathcal{O}(t')\Big\}\,.\label{commut} \eea We
suppressed the momentum index to simplify notation but used the fact
that translational invariance of the bath implies that the
correlation functions are diagonal in momentum. The bath correlation
functions were given in ref.\cite{hobos} (see section 3-B in this
reference) and we just summarize these results: \bea
\mathrm{Tr}_B\rho_B(0) \mathcal{O}(t)\mathcal{O}(t') & = &
\frac{1}{\pi} \int_{-\infty}^{\infty} d\omega~
\mathrm{Im}\widetilde{\Sigma}_{aa}(k;\omega)[1+n(\omega)]~e^{-i\omega(t-t')}
\label{Great}\\\mathrm{Tr}_B\rho_B(0) \mathcal{O}(t')\mathcal{O}(t)
& = & \frac{1}{\pi} \int_{-\infty}^{\infty} d\omega~
\mathrm{Im}\widetilde{\Sigma}_{aa}(k;\omega) ~n(\omega)
~e^{-i\omega(t-t')} \label{Gless} \eea where we used the property
$\mathrm{Im}\widetilde{\Sigma}_{aa}(k;\omega)=-\mathrm{Im}\widetilde{\Sigma}_{aa}(k;-\omega)$\cite{hobos}.
The self energy $\widetilde{\Sigma}$ is obtained from the
discontinuity across the $W-\chi$ lines in the diagram in fig.
(\ref{fig:loop}) and is the same quantity that enters in the
non-equilibrium effective action, and $n(\omega)$ is the equilibrium
distribution function. The active field $\phi_a$ is related to the
fields that create and annihilate the propagating modes in the
medium $\varphi_{1,2}$ as in eq. (\ref{fieldrel}), hence terms of
the form \be \phi_a(t)\phi_a(t') = \cos^2\tm
\varphi_1(t)\varphi_1(t')+\sin^2\tm  \varphi_2(t)\varphi_2(t')+
\frac{1}{2}\sin2\tm \big(
\varphi_1(t)\varphi_2(t')+\varphi_2(t)\varphi_1(t')\big) \,,\ee and
all other   terms in (\ref{commut}) are written accordingly. The
next step requires writing these fields in terms of creation and
annihilation operators in the interaction picture of
$\overline{H}_0$, their expansion is shown in eqn. (\ref{ipfields}).
The resulting products of creation and annihilation operators all
feature phases which are re-arranged to depend separately on the
variable $t$ and $t-t'$, for example \be  a_j  a_j ~e^{2i\omega_j t}
~e^{-i\omega_j (t-t')}~~;~~ a^\dagger_j  a_j
~e^{i\omega_j(t-t')}~~;~~ a^\dagger_i a_j
~e^{i(\omega_i-\omega_j)t}~e^{i\omega_j(t-t')}~~\,,\mathrm{etc.}\,.
\ee The exponentials that depend on $t-t'$, such as $e^{\pm
i\omega_j (t-t')}$ are combined with the exponentials in
(\ref{Great},\ref{Gless}) and the integral in $t'$ in (\ref{commut})
is written as an integral in $\tau=t-t'$. The Wigner-Weisskopf
approximation for the resulting integral yields eqn. (\ref{WW}).
After performing the time integral the terms of the form
$a^\dagger_j a_j$ do not feature any phase, whereas terms of the
form $a_i a_j$ (and their hermitian conjugate) feature terms of the
form $e^{\pm i (\omega_i+\omega_j)t}$, all of these rapidly
oscillating terms average out and are neglected in the ``rotating
wave approximation''\cite{qobooks}, which is tantamount to
time-averaging these rapidly varying terms. The remaining terms can
be gathered together into two different type of contributions,
diagonal and off-diagonal in the $1-2$ indices. The diagonal
contributions do not feature explicit time dependence while the
off-diagonal one features an explicit time dependence of the form
$e^{\pm i(\omega_1-\omega_2) t}$.

\textbf{Diagonal:} The diagonal contributions are  \bea
\label{diagqm}  \frac{d {\rho}_{S,i} }{dt} & = &
\sum_{j=1,2}\sum_{\vk}\Bigg\{ -i \Delta\omega_j(k)
\big[a^\dagger_j(\vk) a_j(\vk),{\rho}_{S,i}(t)\big] -
\frac{\Gamma_j(k)}{2} \Bigg[\Big(1+n(\omega_j(k))\Big)\bigg(
\rho_{S,i}a^\dagger_j(\vk)
a_j(\vk)+a^\dagger_j(\vk) a_j(\vk)\rho_{S,i}\nonumber \\
&-& 2 a_j(\vk)\rho_{S,i}a^\dagger_j(\vk) \bigg)+ n(\omega_j(k))
\bigg( \rho_{S,i}a_j(\vk) a^\dagger_j(\vk)+a_j(\vk)
a^\dagger_j(\vk)\rho_{S,i}-2a^\dagger_j(\vk)\rho_{S,i}a_j(\vk)
\bigg) \Bigg]\Bigg\} \eea where the second order frequency shifts
$\Delta \omega_j(k)$ and the widths $\Gamma_j(k)$ are given in
equations (\ref{omegone}-\ref{delo2}).

\textbf{Off diagonal:} The full expression for the off-diagonal
contributions is lengthy and cumbersome and we just quote the result
for the real part of the quantum master equation, neglecting the
imaginary part which describes a second order shift to the
oscillation frequencies of the off-diagonal coherences.

\bea \label{offdiagqm}\frac{d {\rho}_{S,i} }{dt} & = &
 \sum_{\vk}\Bigg\{
 -\frac{\widetilde{\Gamma}_1(k)}{2}\Bigg[\Big(1+n(\omega_1(k))\Big)
\Big(a^\dagger_2(k;t)a_1(k;t)\rho_{S,i}+\rho_{S,i}a^\dagger_1(k;t)a_2(k;t)-a_2(k;t)\rho_{S,i}a^\dagger_1(k;t) \nonumber \\
 & - & a_1(k;t)\rho_{S,i}a^\dagger_2(k;t)\Big)  + n(\omega_1(k)) \Big(a_2(k;t)a^\dagger_1(k;t)\rho_{S,i}
 +\rho_{S,i}a_1(k;t)a^\dagger_2(k;t)-a^\dagger_2(k;t)\rho_{S,i}a_1(k;t) \nonumber \\
 & - & a^\dagger_1(k;t)\rho_{S,i}a_2(k;t)\Big)\Bigg] -\frac{\widetilde{\Gamma}_2(k)}{2}\Bigg[\Big(1+n(\omega_2(k))
 \Big)\Big(a^\dagger_1(k;t)a_2(k;t)\rho_{S,i}+\rho_{S,i}a^\dagger_2(k;t)a_1(k;t)-a_1(k;t)\rho_{S,i}a^\dagger_2(k;t) \nonumber \\
 & - & a_2(k;t)\rho_{S,i}a^\dagger_1(k;t)\Big)  + n(\omega_2(k)) \Big(a_1(k;t)a^\dagger_2(k;t)\rho_{S,i}
 +\rho_{S,i}a_2(k;t)a^\dagger_1(k;t)-a^\dagger_1(k;t)\rho_{S,i}a_2(k;t) \nonumber \\
 & - & a^\dagger_2(k;t)\rho_{S,i}a_1(k;t)\Big)\Bigg]\Bigg\}\eea where the interaction picture operators
 \be \label{ipas}
 a_j(k;t)= a_j(k;0)~e^{-i\omega_j t} \ee and \be
 \widetilde{\Gamma  }_j(k) =  \frac{1}{2}\sin2\theta_m~
 \frac{\mathrm{Im}\widetilde{\Sigma}_{aa}(k;\omega_j(k))}{\sqrt{\omega_1(k)\omega_2(k)}}
 ~~;j=1,2 \,. \ee

\end{document}